\documentclass[prd,nofootinbib,showpacs,superscriptaddress, preprint]{revtex4-1}
\pdfoutput=1
\usepackage{amsmath,amsfonts,amssymb,graphicx,natbib,bm}
\usepackage{color}
\usepackage{slashed}    
\usepackage[colorlinks,citecolor=blue]{hyperref}
\usepackage{color}
\usepackage{url}
\usepackage{cancel}
\usepackage{slashed}
\usepackage{multirow}
\usepackage{rotating}
\usepackage{braket}
\usepackage{float}
\usepackage{subfigure}

\newcommand{\be}{\begin{eqnarray*}}
\newcommand{\ee}{\end{eqnarray*}}

\newcommand{\bee}{\begin{eqnarray}}
\newcommand{\eee}{\end{eqnarray}}
\newcommand{\beeq}{\begin{equation}}
\newcommand{\eeq}{\end{equation}}

\usepackage{xspace}

\newcommand{\ba}{\begin{array}}
\newcommand{\ea}{\end{array}}
\newcommand{\bd}{\begin{displaymath}}
\newcommand{\ed}{\end{displaymath}}
\newcommand{\besub}{\begin{subequations}}
\newcommand{\eesub}{\end{subequations}}
\newcommand{\bea}{\begin{eqnarray}}
\newcommand{\eea}{\end{eqnarray}}

\def\a{\alpha}

\def\g{\gamma}

\def\l{\lambda}
\def\m{\mu}

\def\G{\Gamma}
\def\D{\Delta}

\def\q2 {q^2}

\def\bt{\begin{table}}
\def\et{\end{table}}

\begin{document}
\title{Two Component Doublet-Triplet Scalar Dark Matter stabilising the Electroweak vacuum  }

\author{Nabarun Chakrabarty}
\email{nabarunc@iitk.ac.in}
\affiliation{Department of Physics, Indian Institute of Technology Kanpur, Kanpur, Uttar Pradesh 208016, India}
\affiliation{Centre for High Energy Physics, Indian Institute of Science, C.V. Raman Avenue, Bangalore 560012, India}
\author{Rishav Roshan}
\email{rishav.roshan@iitg.ac.in}
\author{Arunansu Sil}
\email{asil@iitg.ac.in}
\affiliation{Department of Physics, Indian Institute of Technology Guwahati, Assam-781039, India }

\begin{abstract}  
A two component scalar dark matter scenario comprising an additional scalar doublet and a $Y$ = 0 scalar triplet is proposed.
Key features of the ensuing dark matter phenomenology are highlighted with emphasis on inter-conversion between the two dark matter components. For suitable choices of the model parameters, we show that such inter-conversion can explain the observed relic abundance when the doublet dark matter component has mass in the \emph{desert} region while the triplet component has sub-TeV mass.
This finding is important in the context of such mass regions known to predict under-abundant relic density for the standalone cases of the scalar doublet and triplet. In addition, we also show that the present scenario can stabilise the electroweak vacuum up to the Planck scale in the parameter space responsible for the requisite dark matter observables. 
\end{abstract}


\maketitle
\section{Introduction}

The discovery of a Higgs boson at the Large Hadron Collider (LHC) \cite{Chatrchyan:2012xdj,Aad:2012tfa} completes the particle spectrum of the 
Standard Model (SM). Moreover, the interactions of the discovered boson to fermions and gauge bosons are in good agreement with the SM predictions. Despite this success, certain shortcomings of the SM on both theoretical and experimental fronts keep the hope for additional dynamics
alive and kicking. That the SM alone cannot predict a stable electroweak (EW) vacuum up to the Planck scale for a $t$-quark pole mass at the upper end of its allowed band counts as one such theoretical shortcoming. That is, the SM quartic coupling turns negative while evolving under renormalisation group (RG) thereby destabilising
the vacuum and the energy scale  where that happens is dictated by the $t$-quark mass chosen \cite{Buttazzo:2013uya,Degrassi:2012ry,Tang:2013bz,Ellis:2009tp,EliasMiro:2011aa}. However, augmenting the SM by additional bosonic degrees of freedom (see \cite{EliasMiro:2012ay,Haba:2013lga,Khan:2014kba,Khoze:2014xha,Khan:2015ipa,Gonderinger:2009jp,Gonderinger:2012rd,Chao:2012mx,Gabrielli:2013hma,Chakrabarty:2014aya,Chakraborty:2014oma,Chakrabarty:2015yia,Ghosh:2017fmr,Bhattacharya:2017fid,Garg:2017iva,DuttaBanik:2018emv,Borah:2020nsz} for a partial list) can help the Higgs quartic coupling overcome the destabilising effect coming dominantly
from the $t$-quark. This motivates to look for extensions of the SM scalar sector. On the experimental front, the SM alone cannot postulate a dark matter (DM) candidate, something whose existence is collectively hinted at by observation of galactic rotation curves \cite{Rubin:1970zza}, gravitational lensing \cite{Clowe:2006eq} and anisotropies in the cosmic microwave background. Hitherto the only information known about DM is its relic abundance and is precisely determined by experiments studying anisotropies in cosmic microwave background radiation (CMBR) like Wilkinson Microwave Anisotropy Probe (WMAP)~\cite{Bennett:2012zja} and PLANCK \cite{Aghanim:2018eyx}. Since these experiments do not shed light on DM spin, the possibility that DM can either be a scalar, a fermion or a vector boson, remains open.

Some attractive scalar DM scenarios are based on augmenting the SM scalar sector by additional scalar $SU(2)_L$ multiplets. The most minimal case which in fact transforms trivially under $SU(2)_L$ is that of a scalar singlet (an updated analysis is \cite{gambit}). However, this scenario interacts with the SM only through the Higgs portal and is tightly constrained. The only mass regions where the scalar singlet accounts for all the observed amount of DM are the Higgs resonance dip and the $\sim$ 1 TeV region, with the former being extremely fine-tuned. We therefore focus on the higher $SU(2)_L$ multiplets henceforth that feature gauge interactions.

The next minimal multiplet is an $SU(2)_L$ doublet. This is the popular inert doublet model (IDM) whose neutral CP-even/CP-odd component can serve as a DM candidate \cite{Deshpande:1977rw,LopezHonorez:2006gr, Honorez:2010re, Belyaev:2016lok, Choubey:2017hsq, LopezHonorez:2010tb, Ilnicka:2015jba, Arhrib:2013ela, Cao:2007rm, Lundstrom:2008ai, Gustafsson:2012aj, Kalinowski:2018ylg, Bhardwaj:2019mts}.
Despite the popularity, an aesthetically dissatisfying feature of the IDM is the existence of the $M_{\text{DM}} \in [M_W,500~\text{GeV}]$
region, henceforth called \emph{desert} region which otherwise would have been an interesting range to explore experimentally, wherein only an under-abundant thermal relic density is observed. This underabundance in the desert region is  observed precisely because of the DM annihilating to the SM gauge bosons with a huge annihilation cross-section. However for $M_{\text{DM}}\geq 500$ GeV, small mass splittings between the inert scalars, and, an appropriate value for the DM-Higgs portal interaction, the IDM can indeed predict the observed relic abundance. This can be further traced back to  cancellations that are triggerred between the s-channel, t-channel and contact interaction terms of the DM DM 
$\to V V$ amplitude for such near-degeneracy. And in this fashion, the DM DM $\to V V$ annihilation cross section attains the right value for $M_{\text{DM}}\geq 500$ GeV so as to predict a $\simeq$ 0.1 relic density. There have been efforts in the recent past to revive the IDM desert region by augmenting the IDM with additional fields. Some examples involving additional bosons are~\cite{Bhattacharya:2019fgs} (an additional scalar singlet) and \cite{Borah:2019aeq,Keus:2014jha,Keus:2015xya,Cordero:2016krd,Chakrabarty:2015kmt,Aranda:2019vda,Keus:2019szx} (an additional scalar doublet); while examples involving additional fermions are \cite{Borah:2017dfn,Bhattacharya:2019tqq} (right handed neutrinos). With the scalar singlet and doublet extensions of the IDM already in place, an interesting exercise could be to try extending by a scalar $SU(2)_L$ triplet. And we choose to 
take one with $Y$ = 0 in this study, since it is the most minimal scalar triplet in terms of particle content.

A standalone $Y$ = 0 scalar triplet itself can also be a prospective dark multiplet in the inert limit\cite{Araki:2011hm, Fischer:2011zz, Fischer:2013hwa, Khan:2016sxm,Jangid:2020qgo,Barman:2021ifu,Bell:2020hnr,Bandyopadhyay:2020otm,Ait-Ouazghour:2020slc}. A crucial difference between the inert triplet model (ITM) and
the IDM is that the former features near-degenerate charged and neutral scalars (the mass splitting being 166 MeV only). In addition to cancellations in DM DM $\to V V$, there are also compulsory coannihilations as a fallout of this near-degeneracy. Therefore, for the model to generate the requisite thermal relic abundance and evade the latest direct detection bound, the DM scalar has to at least have $M_{\text{DM}} \simeq 1.8$ TeV. A wider desert region  is thus observed in comparision to the IDM. Therefore, committing to the $Y$=0 scalar triplet scenario annuls the possibility of having of sub-TeV DM. Studies on the stability of the EW vacuum in an IDM and ITM are~\cite{Khan:2015ipa,Khan:2016sxm,Jangid:2020qgo}. We mention here that there is likewise the possibility of populating the triplet desert region by injecting additional fields into the theory~\cite{DuttaBanik:2020jrj}.

Since the quantum number(s) of DM cannot be inferred from fundamental principles or available experimental data, the possibility that DM consists of more than one type of particle
remains alive. Such a notion was first proposed in \cite{Cao:2007fy}
and has in turn spurred many investigations thereafter, a representative 
list being~\cite{Biswas:2013nn,Fischer:2011zz,Bhattacharya:2013hva,Bian:2013wna,Esch:2014jpa,Karam:2015jta,Karam:2016rsz,Bhattacharya:2016ysw,DuttaBanik:2016jzv,Ahmed:2017dbb,Herrero-Garcia:2017vrl,Herrero-Garcia:2018qnz,Poulin:2018kap,Aoki:2018gjf,Bhattacharya:2018cgx,Aoki:2017eqn,Barman:2018esi,Chakraborti:2018aae,Elahi:2019jeo,Borah:2019aeq,Bhattacharya:2019fgs,Biswas:2019ygr,Bhattacharya:2019tqq,Nanda:2019nqy,Maity:2019hre,Khalil:2020syr,Belanger:2020hyh,Nam:2020twn, DuttaBanik:2020jrj, DuttaBanik:2020vfr}. Such multiparticle DM frameworks open up enticing 
DM-DM conversion processes that contribute to the thermal relic abundance but
not to DM-nucleon scattering as looked for at the direct detection (DD) experiments \cite{Akerib:2016vxi,Aprile:2018dbl,Tan:2016zwf,Cui:2017nnn}. A multipartite DM scenario therefore can evade the ever tightening bound on the DD rates while enlarging the parameter space compatible with the observed relic density. We propose one such multi-component DM framework in this work that combines the two single component scenarios discussed above, i.e, an inert scalar doublet and a $Y$ = 0 scalar triplet. The key takeaway from the discussion on the IDM and ITM is that certain mass regions in both cases predict underabundant relic densities precisely due to the gauge interaction-mediated co(annihilations). We aim to investigate here if DM-DM conversion can revive these mass regions in the proposed multi-component setup. In other words, the primary motive of this study is to revive the DM mass regions corresponding to the two $SU(2)_L$ multiplets that are forbidden in the respective standalone cases. Since these inert multiplets interact with the SM-like doublet via the scalar potential, they tend to aid to EW vacuum stability by generating positive contributions to the RG running of the SM-like quartic coupling. Therefore, we also wish to find out if the parameter space compatible with DM relic density and DD can stabilise the EW vacuum till the Planck scale.   

This study is organised as follows. The model is detailed in section~\ref{Model} and the relevant constraints are discussed in section~\ref{constraints}. Section~\ref{DMPH} throws light on the multi-component DM phenomenology with an emphasis on DM-DM conversion. Section~\ref{EWSB} discusses EW vacuum (meta)stability and comments on the results obtained upon demanding the DM constraints and a (meta)stable vacuum 
up to the Planck scale together. We conclude in section~\ref{Conclusions}. Important formulae are relegated to the Appendix.

\section{Model}
\label{Model}

In the present study, we extend the SM by an $SU(2)_L$ scalar doublet $\Phi$ with $Y = \frac{1}{2}$ and a hyperchargeless $SU(2)_L$ scalar triplet $\mathcal{T}$. A discrete symmetry $Z_{2}\times Z_{2}^{\prime}$ is introduced under which the SM fields are trivial while the additional scalar multiplets are charged. We provide in Table \ref{t1} the quantum numbers of all the scalars in the scenario under both gauge and discrete symmetries. The 
$Z_{2}\times Z_{2}^{\prime}$ ensures stability of the newly introduced scalar multiplets as a result of which
the neutral scalars, if lightest within their respective multiplets, can be DM candidates. Therefore the present setup can accommodate a two component DM scenario. We add that $\Phi$ ($\mathcal{T}$) hereafter can be referred to  
as \emph{inert} doublet (triplet) since it does not pick up a vacuum expectation value (VEV) by virtue of the discrete symmetry. 
\begin{table}[H]
\begin{center}
\vskip 0.5 cm
\begin{tabular}{|c|c|c|c|c|}
\hline
    Particle & $SU(2)$ &   $U(1)_Y$  & $Z_2$ &  $Z_2^{\prime}$           \\
\hline        
$H$         &  2     &  $\frac{1}{2}$         &  + & +                \\
\hline
$\Phi$      &  2     & $\frac{1}{2}$         &  + & -                \\
\hline
$\mathcal{T}$         &  3     &  0         &  - & +                \\

\hline
\end{tabular}
\end{center}  
\caption{Quantum numbers of the SM Higgs doublet $H$ and the inert multiplets $\Phi$ and 
$\mathcal{T}$.}
\label{t1}
\end{table}

The most general renormalisable scalar potential consistent with $SU(2)_{L}\times U(1)_Y\times Z_2 
\times Z_{2}^{\prime}$ for the given scalar content, $V(H,\Phi,\mathcal{T})$, consists of (i) $V_H$: terms involving $H$ alone, (ii) $V_{\Phi}$ : terms involving $\Phi$ alone, (iii) $V_{\mathcal{T}}$: terms involving 
$\mathcal{T}$ alone and (iv) $V_{\rm{int}}$:  interactions involving all $H,~\Phi,~\mathcal{T}$. That is, 
\begin{eqnarray}
V(H,\Phi,\mathcal{T}) &=& V_H + V_{\Phi} + V_{\mathcal{T}}  + V_{\text{int}} .
\label{p1}
\end{eqnarray}
where
\besub
\bea
V_{H} &=& -\mu_H^2 H^\dag H + \lambda_H (H^\dag H)^2,  \\
V_{\Phi} &=& \mu_{\Phi}^2  \Phi^\dag \Phi + \lambda_{\Phi} (\Phi^\dag \Phi)^2,  \\
V_{\mathcal{T}} &=& \frac{M_{T}^2}{2}~{\rm{Tr}}[\mathcal{T}^2]+\frac{\lambda_T}{4!}(~{\rm{Tr}}[\mathcal{T}^2])^2,  
\eea
\eesub
and 
\bea
V_{\rm{int}}&=& 
  \lambda_{1}
(H^\dagger H)(\Phi^\dagger \Phi)+ 
\lambda_2 (\Phi^\dagger H)(H^\dagger \Phi)+ \frac{1}{2} \lambda_3
[(\Phi^\dagger H)^2 + (H^\dagger \Phi)^2]+\frac{\lambda_{HT}}{2}(H^\dag H)~{\rm{Tr}}[\mathcal{T}^2]\nonumber \\ &+& \frac{\l_{\Phi \mathcal{T}}}{2}(\Phi^\dagger \Phi)~{\rm{Tr}}[\mathcal{T}^2].
 \label{p1int}
 \eea
Following electroweak symmetry breaking (EWSB), the CP-even neutral component of 
$H$ obtains a VEV $v=246$ GeV. On the other hand, the $Z_2$ and $Z_2^{\prime}$ ensures that the neutral components of $\Phi$ and $\mathcal{T}$ do not pick up VEVs, as stated before. The scalar fields can then be parameterised as
\bea
H = \left( \begin{array}{c}
                          0  \\
                          \frac{1}{\sqrt{2}}(v+h)  
                 \end{array}  \right) \, ,                     
                 ~~~~~\Phi =\left( \begin{array}{c}
                          H^+  \\
                          \frac{1}{\sqrt{2}}(H_0+iA_0)  
                 \end{array}  \right), 
 \mathcal{T} =\left( \begin{array}{c}\
                           \frac{1}{\sqrt{2}}T_0~~~~-T^+ \\
        				      -T^- ~~~~-\frac{1}{\sqrt{2}}T_0
                 \end{array}  \right), \
\eea
and after the EWSB, the masses of the physical scalars are given by 
\bea
m_h^2&=&2 \lambda_H v^2,\nonumber  \\
m^{2}_{H^{\pm}}&=&\mu_{\Phi}^{2}+\lambda_{1}\frac{v^{2}}{2}, \nonumber \\
m_{H_0}^{2}&=&\mu_{\Phi}^{2}+(\lambda_{1}+\lambda_{2}+\lambda_{3})\frac{v^{2}}{2}, \nonumber \\
m_{A_0}^{2}&=&\mu_{\Phi}^{2}+(\lambda_{1}+\lambda_{2}-\lambda_{3})\frac{v^{2}}{2},\nonumber \\
m_{T_0,T^\pm}^{2}&=& M_T^2+ \frac{\lambda_{HT}}{2}v^2 .
\label{mass}   
\eea 

In Eq.(\ref{mass}), $m_h=125.09$ GeV \cite{deFlorian:2016spz}, is the mass of SM Higgs. The fact that the scalars from the $\Phi$ multiplet have different masses at the tree level itself paves way for the possibility that $H_0$ is the lightest and is therefore a DM candidate. Unlike $\Phi$, the charged and neutral members of $\mathcal{T}$ have degenerate masses at the tree level. However, thanks to radiative effects, this degeneracy is lifted at the one-loop level leading to the following mass-splitting.
\bea
\Delta m&=& (m_{T^{\pm}}-m_{T_0})_{1-\text{loop}} = \frac{\alpha ~m_{T_0}}{4\pi}\bigg{[}f\bigg{(}\frac{M_W}{m_{T_0}}\bigg{)}\nonumber \\ &-&c_{W}^2f\bigg{(}\frac{M_Z}{m_{T_0}}\bigg{)}\bigg{]},
\label{loop1}   
\eea 
where $\a$ is the fine structure constant, $M_W,~M_Z$ are the masses of the W and Z bosons, $c_W=\cos\theta_W=M_W/ M_Z$ and $f(x)=-\frac{x}{4}\bigg{[}2x^3\rm{ln}(x)+(x^2-4)^{3/2}\rm{ln}\bigg{(}\frac{x^2-2-x\sqrt{x^2-4}}{2}\bigg{)}\bigg{]}$where $x=\frac{M_{W,Z}}{m_{T_0}}$. It turns out that 
in the limit $x\rightarrow0$ $i.e.$ $m_{T_0} \gg$ $M_W$ or $M_Z$, $f(x)\rightarrow 2\pi x$ and $\Delta m$ can be expressed as \cite{Sher:1995tc,Cirelli:2005uq}
\bea
\Delta m&=&\frac{\a}{2}M_W\sin^2\frac{\theta_W}{2} = 166~ \rm{MeV}. 
\label{deltam}   
\eea 
One now gathers that $T_0$ is also stable and can be a DM candidate. 

We now turn to identify the independent parameters in this scenario. A counting of parameters in the Lagrangian yields $\{\mu_H,\mu_\Phi,M_T,\l_H,\l_\Phi,\l_T,\l_1,\l_2,\l_3,\l_{HT},\l_{\Phi T} \}$: 11 parameters. Now, 
$\mu_H$ is eliminated demanding that the Higgs tadpole vanishes. At this level, we invoke 
$\l_L \equiv \frac{\l_1 + \l_2 + \l_3}{2}$ as an independent parameter that quantifies the $H_0-H_0-h$ portal interaction and hence is of profound importance in a Higgs portal DM scenario such as the IDM. Similarly, 
$\l_{HT}$ parameterises the strength of the $T_0-T_0-h$ portal coupling and is treated as an independent parameter
henceforth. So is $\l_{\Phi T}$ since it parameterises the 
$T_0 T_0 \to H_0 H_0$ conversion amplitude. Finally, 
$\mu_\Phi,M_T,\l_1,\l_2$ and $\l_3$ can be traded off with the inert masses and $\l_L$ using
\besub
\bea
\mu^2_\Phi &=& m^2_{H_0} - \l_L v^2, \\
M^2_T &=& m^2_{T_0} - \frac{\l_{HT}}{2}v^2, \\
\l_1 &=& 2 \l_L + \frac{2(m^2_{H^+} - m^2_{H_0})}{v^2}, \label{l1}\\
\l_2 &=& \frac{m^2_{H_0} + m^2_{A_0} - 2 m^2_{H^+}}{v^2}, \label{l2}\\
\l_3 &=& \frac{m^2_{H_0} - m^2_{A_0}}{v^2}\label{l3}.
\eea
\eesub
The independent parameters in the scalar sector are therefore
$$\{m_h,m_{H_0},m_{A_0},m_{H^{\pm}},m_{T_0},~\lambda_{L},~\lambda_{HT},\l_{\Phi T},\l_H,\l_\phi,\l_T\}.$$
 In passing, we add that $\l_\Phi$ and $\l_T$ parameterise self-interaction within their respective inert sectors and hence are not phenomenologically that much relevant for the ensuing DM analysis. However, these couplings can indeed be constrained from the theoretical requirements of perturbativity, unitarity and positivity of the scalar potential. A detailed discussion can be seen in Section~\ref{constraints}.

\section{Constraints}
\label{constraints}
The present scenario is subject to the following theoretical and experimental constraints.

\subsection{Theoretical constraints: Perturbativity, positivity of the scalar potential and unitarity}
The present model is deemed perturbative if the scalar quartic couplings obey $|\l_i| \leq 4\pi$. Further, the gauge and Yukawa couplings must also satisfy $|g_i|,~|y_i|\leq \sqrt{4\pi}$. 

The introduction of additional scalars opens up additional directions in field space. The following conditions ensure that the potential remains bounded from below (BFB) along each such direction
\besub
\bea
\lambda_H ,~ \lambda_T ,~ \lambda_{\Phi} \geq 0,  \\
\lambda_{HT}~+~\sqrt{\frac{2}{3}\lambda_{H}\lambda_{T}}~\geq 0,  \\
\lambda_{1}~+~2\sqrt{\lambda_{H}\lambda_{\Phi}}~\geq 0,  \\
\lambda_{1}~+\lambda_{2}~-~|\l_3|~+2\sqrt{\lambda_{H}\lambda_{\Phi}}~\geq 0,  \\
\lambda_{\Phi T}~+~\sqrt{\frac{2}{3}\lambda_{\Phi}\lambda_{T}}~\geq 0. 
\,\,  
\eea
\label{copos} 
\eesub
Additional restrictions on the scalar potential come from unitarity. For this model, a couple of $2 \to 2$ scattering matrices can be constructed in the basis of neutral and singly charged two-particle states. Unitarity demands that the absolute value of each eigenvalue of
these scattering matrices must be $\leq 8\pi$. An
element of the scattering matrix is proportional to a quartic coupling in the high energy limit. Following this prescription, one derives for this model
\bea
|\l_1 \pm \l_2| \leq 8\pi;~~~|\l_1 \pm \l_3| \leq 8\pi
;~~~|\l_1 +2\l_2 \pm \l_3| \leq 8\pi;~~~|\l_T| \leq 24\pi;\nonumber\\
|\l_{\Phi T}| \leq 8\pi;~~ |\l_{HT}| \leq 8\pi; 
~~~|\l_H + \l_\Phi \pm \sqrt{ (\l_H - \l_\Phi)^2 + \l_2^2}| \leq  8\pi; \nonumber\\
|\l_H + \l_\Phi \pm \sqrt{ (\l_H - \l_\Phi)^2 + \l_3^2}| \leq  8\pi.
\eea
Further, since the present study discusses Higgs vacuum stability, the conditions of perturbativity, positivity of the scalar potential and unitarity have to be met at each intermediate scale while evolving from the EW scale to a higher scale under RG.

\subsection{DM observables}
For the present scenario to be a successful DM model, the thermal relic abundance it predicts must lie in the observed band. Adopting the latest result from the measurement of relic abundance by the Planck experiment, we demand from our model
\bea
0.119 \lesssim \Omega h^2 < 0.121.
\eea
Non-observation of DM-nucleon scattering at the terrestrial experiments have put upper limits on the corresponding cross section as a function of DM mass.
We abide by in our study the bound on the spin-independent direct detection (SI-DD) cross section from XENON-1T, the experiment predicting the most stringent bound.

\subsection{LHC diphoton signal strength} 
The dominant amplitude for the $h \to \gamma \gamma$ in the SM reads
\besub
\bea
\mathcal{M}^{\text{SM}}_{h \to \gamma \gamma} &=& 
\frac{4}{3} A_{1/2}\Big(\frac{M^2_h}{4 M^2_t}\Big)
+ A_1\Big(\frac{M^2_h}{4 M^2_W}\Big).
\label{htogaga_SM}
\eea
\eesub
We have neglected the small effect of fermions other than the $t$-quark in Eq.(\ref{htogaga_SM}). The presence of additional charged scalars in the present framework implies modification to the $h \to \gamma \gamma$ amplitude \emph{w.r.t.} the SM. This additional scalar contribution reads
\bea
M^S_{h \to \gamma \gamma} &=& \frac{\l_{h H^+ H^-}v}
{2 m^2_{H^+}} A_0\bigg(\frac{m^2_h}{4 m^2_{H^+}}\bigg)
+ \frac{\l_{h T^+ T^-}v}
{2 m^2_{T^+}} A_0\bigg(\frac{m^2_h}{4 m^2_{T^+}}\bigg)
\eea 
where
\besub
\bea
\l_{h H^+ H^-} &=& \l_1 v, \\
\l_{h T^+ T^-} &=& \l_{HT} v.
\eea
\eesub

The total amplitude and the decay width then become
\bea
\mathcal{M}_{h \to \gamma \gamma} &=& 
\mathcal{M}^{\text{SM}}_{h \to \gamma \gamma} +
\mathcal{M}^{\text{S}}_{h \to \gamma \gamma}, \\
\Gamma_{h \to \gamma \gamma} &=& \frac{G_F \a^2 m_h^3}{128 \sqrt{2} \pi^3} |\mathcal{M}_{h \to \gamma \gamma}|^2.\eea
where $G_F$ is the Fermi constant. The various loop functions are listed below~\cite{Djouadi:2005gj}.
\besub
\bea
A_{1/2}(x) &=& \frac{2}{x^2}\big((x + (x -1)f(x)\big), \\
A_1(x) &=& -\frac{1}{x^2}\big((2 x^2 + 3 x + 3(2 x -1)f(x)\big), \\
A_0(x) &=& -\frac{1}{x^2}\big(x - f(x)\big),  \\
\text{with} ~~f(x) &=& \text{arcsin}^2(\sqrt{x}); ~~~x \leq 1 
\nonumber \\
&&
= -\frac{1}{4}\Bigg[\text{log}\frac{1+\sqrt{1 - x^{-1}}}{1-\sqrt{1 - x^{-1}}} -i\pi\Bigg]^2;\nonumber \\&&x > 1.
\eea
\eesub
where $A_{1/2}(x), A_1(x)$ and 
$A_0(x)$ are loop functions corresponding to spin-$\frac{1}{2}$, spin-1 and spin-0 particles in the loop.
The signal strength for the $\gamma\gamma$ channel is defined as
\bea
\mu_{\gamma\gamma} &=&  \frac{\sigma(pp \to h)\text{BR}(h \to \gamma \gamma)}{\Big[\sigma(pp \to h)\text{BR}(h \to \gamma \gamma)\Big]_{\text{SM}}}.
\eea
Given the inert multiplets do not modify the $pp\to h$ production,
\bea
\mu_{\gamma\gamma} &=&  \frac{\text{BR}(h \to \gamma \gamma)}{\Big[\text{BR}(h \to \gamma \gamma)\Big]_{\text{SM}}}, \\
& \simeq & \frac{\Gamma^{\text{SM}}_{h\to\gamma\gamma}}{\Gamma_{h\to\gamma\gamma}}.
\eea 
In order to ensure that $\mu_{\g\g}$ lies within the experimental uncertainties, the analysis should respect the latest signal 
strength from ATLAS \cite{Aaboud:2018xdt} and CMS \cite{Sirunyan:2018koj}. The measured value of $\mu_{\g\g}$ are given by $\mu_{\g\g}=0.99\pm 0.14$ from ATLAS and $\mu_{\g\g}=1.17\pm 0.10$ from CMS. The constraints have been imposed at 2$\sigma$.

\subsection{Disappearing charged track}

The smallness of the mass-splitting between
$T^+$ and $T_0$ implies that the 
$T^+ \to T_0 \pi^+$ decay width is tiny. The parent particle $T^+$ will therefore travel a macroscopic distance ($\mathcal{O}$(1cm))
before decaying thereby leading to multiple hits in the LHC tracking devices. In addition, the $\pi^+$ would be too \emph{soft} to be detected. As a result, a \emph{disappearing charged track} (DCT) signal would be seen. The lighter is $T^+$, the higher is the $pp \to T^+ T^-$ production cross section at the LHC and hence, the higher would be the number of DCT events. It then follows that non-observation of such events at the LHC would lead to lower bounds on the mass of $T^+$. It was shown recently \cite{Chiang:2020rcv} that non-observation of DCT signals at the 13 TeV, integrated luminosity (L) = 36 fb$^{-1}$ excludes a real triplet scalar lighter than 275 GeV. The reach can extend to 590 GeV and 745 GeV with 
L = 300 fb$^{-1}$ and 3000 fb$^{-1}$ respectively. We have therefore maintained $m_{T^+} >$ 275 GeV throughout the analysis in light of the current constraint.

\subsection{Oblique parameters}
The additional scalars present in this setup can induce potentially important contributions to the oblique ($S,T,U$) parameters That is, for $X=S,T,U$, one can write
\bea
\Delta X &=& \Delta X_{\text{ID}} + \Delta X_{\text{IT}},
\eea
where the subscripts ID (IT) denotes the contribution from the inert doublet (triplet). We have for the inert doublet,
\besub
\bea
\D S_{\text{ID}} &=& \frac{1}{2\pi}\bigg[\frac{1}{6}\ln\bigg(\frac{m_{H_0}^2}{m_{H^\pm}^2}\bigg)-\frac{5}{36}+\frac{m_{H_0}^2m_{A_0}^2}{3(m_{A_0}^2-m_{H_0}^2)^2}+\frac{m_{A_0}^4(m_{A_0}^2-3m_{H_0}^2)}{6(m_{A_0}^2-m_{H_0}^2)^3}\ln\bigg(\frac{m_{A_0}^2}{m_{H_0}^2}\bigg)\bigg], \\ \nonumber \\
\Delta T_{\text{ID}} &=& \frac{1}{16 \pi s^2_W M_W^2}\Big[F(m^2_{H^+},m^2_{H_0}) + F(m^2_{H^+},m^2_{A_0}) 
- F(m^2_{H_0},m^2_{A_0})\Big] \label{ID_T}, \\ \nonumber \\
\D U_{\text{ID}} &=& 0. 
\eea
\eesub 
In the above,
\bea
F(x,y) &=&  \frac{x+y}{2} - \frac{xy}{x-y}~{\rm ln} \bigg(\frac{x}{y}\bigg)~~~ {\rm for} ~~~x \neq y \,, \nonumber \\
&=& 0~~~ {\rm for} ~~~ x = y.
\eea
While for the inert triplet,
\besub
\bea
\D S_{\text{IT}} &=& 0, \\
\D T_{\text{IT}} &=& \frac{1}{16 \pi s^2_W M_W^2}
 F(m^2_{T^+},m^2_{T_0}) \label{IT_T}, \\
&\simeq& 
\frac{(\Delta m)^2}{24 \pi s^2_W M_W^2}~~~~~\text{since $\Delta m < < m_{T_0}$}, \\
\D U_{\text{IT}} &=& -\frac{1}{3\pi}\bigg{[}m_{T_0}^4\rm{ln}\bigg{(}\frac{m_{\it{T}_{\rm 0}}^2}{m_{\it{T^{\pm}}}^2}\bigg{)}\frac{(3m_{\it{T^{\pm}}}^2-m_{\it{T}_{\rm 0}}^2)}{(m_{\it{T}_{\rm 0}}^2-m_{\it{T^{\pm}}}^2)^3}+\frac{5(m_{\it{T}_{\rm 0}}^4+m_{\it{T^{\pm}}}^4)-22m_{\it{T}_{\rm 0}}^2m_{\it{T^{\pm}}}^2}{6(m_{\it{T}_{\rm 0}}^2-m_{\it{T^{\pm}}}^2)^2}\bigg{]}\\ \nonumber 
&\simeq & \frac{\Delta m}{3\pi ~m_{T^{\pm}}}
\eea
\eesub
The most updated bounds read~\cite{Zyla:2020zbs}
\bea
\Delta S = 0.02 \pm 0.10,~~\Delta T = 0.07 \pm 0.12,~~\Delta U = 0.00 \pm 0.09.
\eea
The stated bounds have been imposed at 2$\sigma$ in our analysis. A few comments are in order. First, the contribution of an inert real scalar triplet to the oblique parameters is found to be at most $\mathcal{O}$ $\big(\frac{\Delta m}{m_{T_0}}\big)$ or further suppressed. The contributions are therefore negligible in comparison to the corresponding ones from the inert doublet. Secondly, a non-zero 
$T$-parameter indicates that the $\rho$-parameter deviates from unity at one-loop (since the doublets and the VEVless triplet present in this model predict $\rho$ = 1 at tree level.). However, ensuring that $T$ stays within the stipulated bound is tantamount to obeying the $\rho$-parameter constraint.

\section{Dark Matter phenomenology}
\label{DMPH}

\begin{figure}[]
\centering
\subfigure[]{
\includegraphics[scale=0.35]{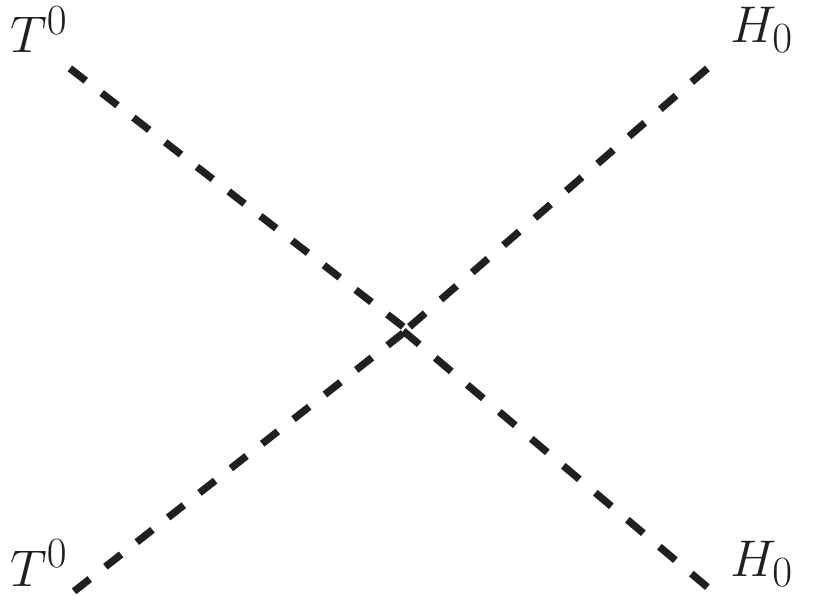}}
\subfigure[]{
\includegraphics[scale=0.35]{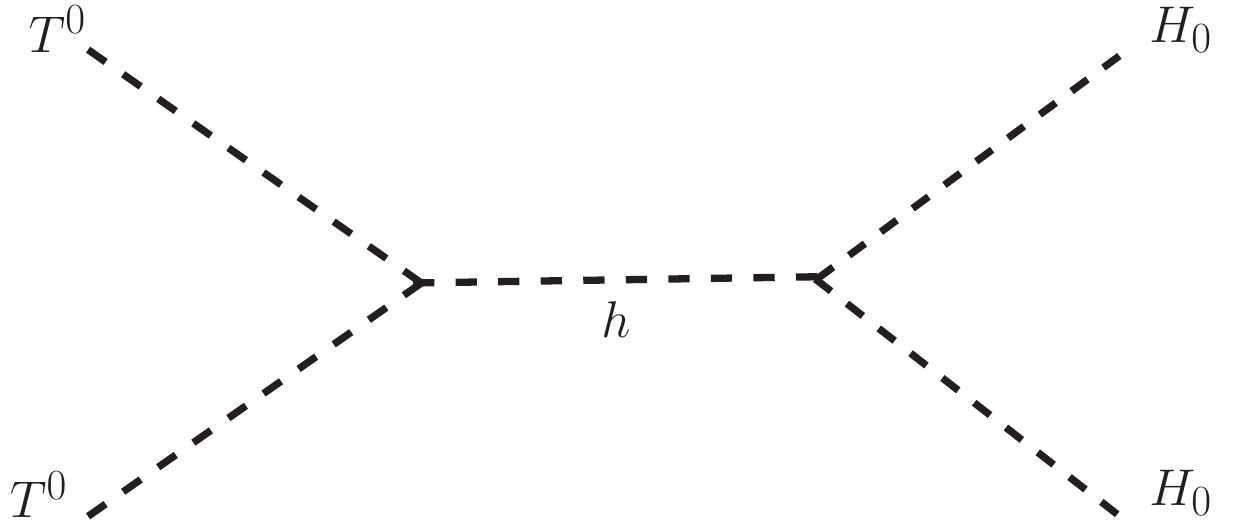}}
\subfigure[]{
\includegraphics[scale=0.35]{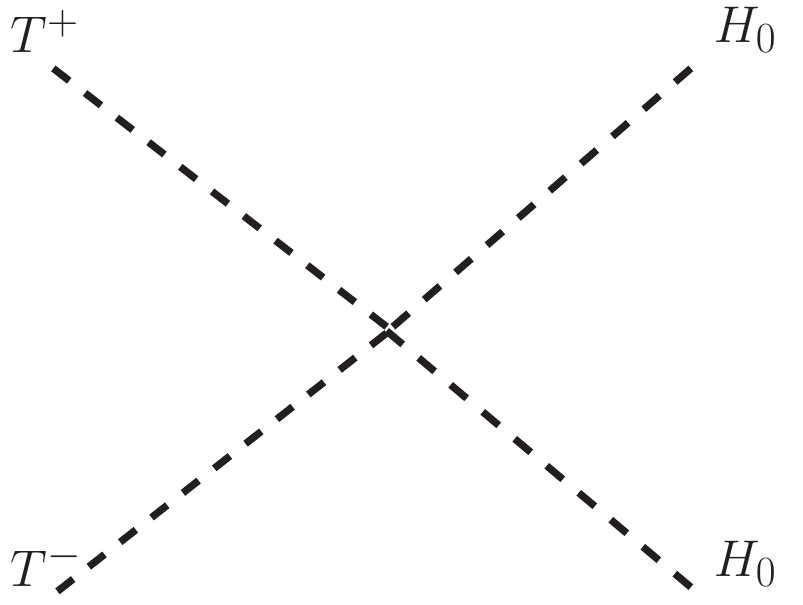}}
\subfigure[]{
\includegraphics[scale=0.35]{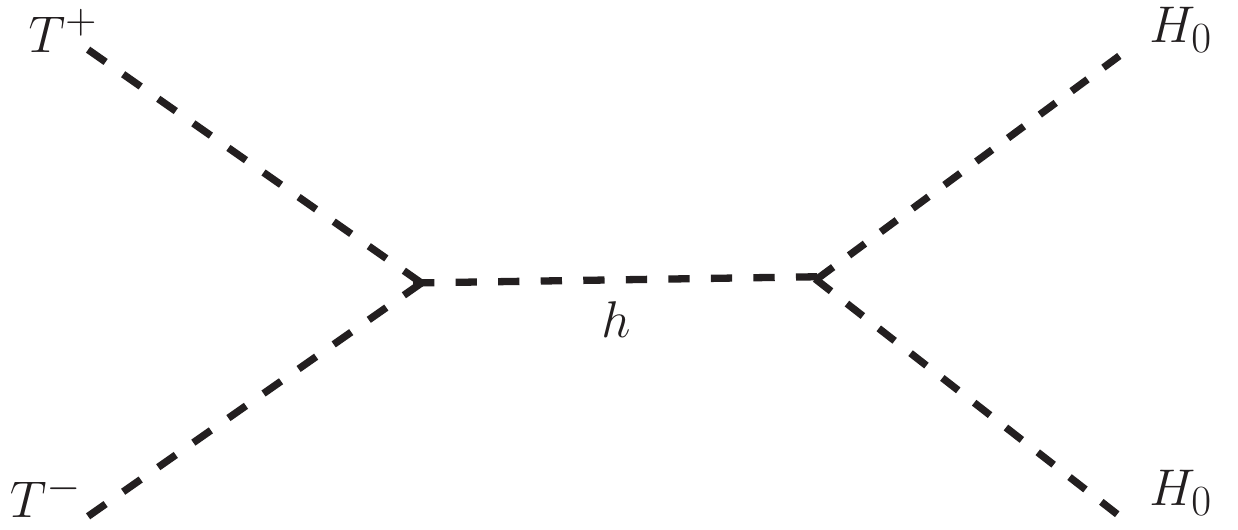}}
\caption{DM-DM conversion channels, assuming $m_{T_0}>m_{H_0}$.}
\label{feynconv}
\end{figure}

The present setup being a two component DM scenario, has the discrete symmetry $Z_2\times Z_{2}^{\prime}$ that remain unbroken throughout and guarantee the stability of the DM candidates. To find the individual contributions to the relic density, one requires to evaluate the yields of both the DM species by solving coupled Boltzmann equations. In order to do so, first we identify all the relevant annihilation and coannihilation channels of both, (we refer the reader \cite{Bhattacharya:2019tqq} and \cite{DuttaBanik:2020jrj} where all such channels for IDM and ITM are listed respectively). Apart from these, diagrams responsible for DM-DM conversion are shown in Fig.\ref{feynconv} assuming $m_{T_0} > m_{H_0}$. This inter-conversion of the two DM particles plays a significant role in our analysis and in turn, makes the set of Boltzmann equations coupled. Next to calculate the DM relic abundance, we implement the model file in LanHEP \cite{Semenov:2014rea} and then pass the files generated in the LanHEP to the micrOMEGAS \cite{Belanger:2014vza}. For inspecting the current scenario, we consider the all constraints discussed 
in section \ref{constraints}.


\subsection{Relic Density}
In order to obtain the comoving number densities of the both the DM particles, one needs to solve a set of coupled Boltzmann equation in the present setup as the conversion of one DM to another plays a non-trivial role. Due to the involvement of two DM particles, it is always better to redefine the usual $x$ parameter from 
$\frac{m_{\rm{DM}}}{\tilde{T}}$ to 
$\frac{\mu_{\text{DM}}}{\tilde{T}}$, where $\mu_{\text{DM}}$ is nothing but the reduced mass   expressed as: $~\mu_{\text{DM}}= \frac{m_{H_0}m_{T_0}}{m_{H_0}+m_{T_0}}$ whereas $\tilde{T}$ represents the temperature of the Universe. Finally, we write the coupled Boltzmann equations in terms of parameter $x=\frac{\m_{\text{DM}}}{\tilde{T}}$  and the co-moving number density $Y_{\rm DM} = n_{\rm DM}/s$ ($s$ being the entropy density) redefined as $y_i$ $i = H_0,T_{0}$) via $y_i=0.264M_{Pl}\sqrt{g_*}\mu Y_{i}$, as below\footnote{We use the notation from a recent article on two component DM  \cite{Bhattacharya:2018cgx}.}
\besub
\bea
\frac{dy_{H_0}}{dx} &=& \frac{-1}{x^2}\bigg{[}\langle \sigma v_{H_0H_0\rightarrow XX}\rangle \left(y_{H_0}^{2}-(y_{H_0}^{EQ})^2\right)~+~\langle \sigma v_{H_0H_0\rightarrow T_{0}T_{0}}\rangle \left ( y_{H_0}^{2}-\frac{(y_{H_0}^{EQ})^2}{(y_{T_{0}}^{EQ})^2} y_{T_{0}}^{2}\right)\Theta(m_{H_0}-m_{T_{0}})\nonumber \\
&& 
-~\langle\sigma v_{T_{0}T_{0} \rightarrow H_0 H_0}\rangle \left( y_{T_{0}}^{2}-\frac{(y_{T_{0}}^{EQ})^2}{(y_{H_0}^{EQ})^2}y_{H_0}^{2}\right)~\Theta(m_{T_{0}}-m_{H_0})\bigg{]}  , \\
\frac{dy_{T_{0}}}{dx} &=& \frac{-1}{x^2}\bigg{[} \langle \sigma v_{T_{0}T_{0}\rightarrow XX} \rangle \left (y_{T_{0}}^{2}-(y_{T_{0}}^{EQ})^2\right )~+~
\langle \sigma v_{T_{0}T_{0}\rightarrow H_0 H_0}\rangle \left (y_{T_{0}}^{2}-\frac{(y_{T_{0}}^{EQ})^2}{(y_{H_0}^{EQ})^2}y_{H_0}^{2}\right )\Theta(m_{T_{0}}-m_{H_0}) \nonumber \\
&&
-~\langle \sigma v_{H_0H_0 \rightarrow T_{0}T_{0}}\rangle \left (y_{H_0}^{2}-\frac{(y_{H_0}^{EQ})^2}{(y_{T_{0}}^{EQ})^2}y_{T_{0}}^{2}\right )
\Theta(m_{H_0}-m_{T_{0}})\bigg{]}.
\eea
\label{BE}
\eesub

\noindent One can relate $y_i^{EQ}$ in the same way as $y_{i}$ via $y_i^{EQ}= 0.264M_{Pl}\sqrt{g_*}\mu_{\text{DM}} Y_{i}^{EQ}$,  where $Y_{i}^{EQ}$ is the equilibrium density $Y_{i}^{EQ}$ defined in terms of $\mu_{\text{DM}}$ as
\bea
Y_{i}^{EQ}(x) = 0.145\frac{g}{g_*}x^{3/2}\bigg{(}\frac{m_{i}}{\mu_{\text{DM}}}\bigg{)}^{3/2}e^{-x\big{(}\frac{m_{i}}{\mu_{\text{DM}}}\big{)}}.
\eea 
Here $M_{\rm Pl} = 1.22\times10^{19} ~{\rm GeV}$,  $g_{*}=106.7$, $m_i=m_{H_0},m_{T_0}$, $X$ represents all the SM particles, $H^{\pm},A_0$ and $T^{\pm}$, and finally, the thermally averaged effective annihilation cross-section inclusive of both the annihilation and DM-DM conversion processes can be expressed as
\bea
\langle \sigma v\rangle = \frac{1}{8m^{4}_{i}\tilde{T} K_2^2(\frac{m_{i}}{T})}\int\limits^{\infty}_{4m_{i}^2}\sigma(s-4m_{i}^2)\sqrt{s}K_1\bigg{(}\frac{\sqrt{s}}{\tilde{T}}\bigg{)}ds,
\label{eq:sigmav}
\eea
and is evaluated at $\tilde{T}_f$. The freeze-out temperature $\tilde{T}_f$ can be derived by equating the DM interaction rate $\Gamma = n_{\rm DM} \langle \sigma v \rangle$ with the  expansion rate of the universe $H(\tilde{T}) \simeq \sqrt{\frac{\pi^2 g_*}{90}}\frac{T^2}{M_{\rm Pl}}$. In  Eq.(\ref{eq:sigmav}), $K_{1,2}(x)$ represents the modified Bessel functions. Finally, in Eq.(\ref{BE}), $\Theta$ function is used in order to explain the conversion process (corresponding to Fig.\ref{feynconv}) of one DM to another which strictly depends on the mass hierarchy of DM particles. 

At this stage, it is perhaps pertinent to mention that $H^{\pm}$ (heavier than $H_0$) and $T^{\pm}$ (heavier than $T_0$) 
are expected to be in equilibrium with the thermal plasma by virtue of their electromagnetic interactions as well as their interactions with the Higgs.  Apart from being in equilibrium, the $H^+$ can decay into $H_0l^+\bar{\nu}$ and $T^+$ can decay to $T_0\pi^+$ via off-shell $W$-bosons. Finally $A_0$ being heavier than $H_0$ can always decay to $H_0$ and the SM fermions via an off shell $Z$. The heavier scalars within the dark multiplets are thus not cosmologically stable. These coupled equations now have to be solved numerically to find the asymptotic abundance of the DM particles, $y_{i} \left (\frac{\mu_{\text{DM}}}{m_{i}}x_{\infty} \right)$, which can be further used to calculate the relic density:
\bea
\Omega_{i}h^2 &=& \frac{854.45\times 10^{-13}}{\sqrt{g_{*}}}\frac{m_{i}}{\mu_{\text{DM}}}y_{i}\left ( \frac{\mu_{\text{DM}}}{m_{i}}x_{\infty}\right ),
\eea
where $x_{\infty}$ indicates a very large value of $x$ after decoupling. Total DM relic abundance is then given as
\bea
\Omega_{\rm Total} {h}^2 =\Omega_{T_0} {h}^2 + 
\Omega_{H_0} {h}^2\,\, .
\label{totalrelic}
\eea
It is to be noted that total relic abundance must satisfy the DM relic density
obtained from Planck \cite{Aghanim:2018eyx}
\bea
\Omega_{\rm Total} {h}^2 =0.1199{\pm 0.0027}\, .
\label{totalrelic_val}
\eea

\subsection{Direct and Indirect detection}
Experiments like LUX \cite{Akerib:2016vxi}, PandaX-II \cite{Tan:2016zwf,Cui:2017nnn} and Xenon-1T \cite{Aprile:2017iyp,Aprile:2018dbl} look for signals of DM-nucleon scattering. And, non-observation of the same have led to upper bounds on the DM-nucleon scattering cross-section as a function of DM mass. It must be added that, in principle, inelastic direct detection scatterings can also get triggered in case the mass gap between the DM and the next heavier particle within the multiplet
is below $\sim$ 150 keV \cite{Arina:2009um}. Being a two component DM scenario, in the present model both the DM particles would appear in direct search experiments.
However, one should take into account the fact that direct detection
cross sections of both are to be rescaled by the corresponding relic density fractions. Hence, the effective direct detection cross-section of triplet scalar DM $T_0$ is given as \cite{Ayazi:2015mva}
\bea
\sigma_{\rm {T_0,eff}}= \frac{\Omega_{T_0}}{\Omega_{\rm Total}}\frac{\lambda_{HT}^2}{4\pi}\frac{1}{m_h^4} f^2
\frac{m_N^4}{(m_{T_0}+m_N)^2},
\label{tripletdd}
\eea
and similarly the effective direct detection cross-section of $H_0$ is expressed as \cite{Borah:2019aeq}
\bea
\sigma_{\rm {H_0,eff}}=\frac{\Omega_{H_0}}{\Omega_{\rm Total}}\frac{\lambda_{L}^2}{4\pi}\frac{1}{m_h^4} f^2
\frac{m_N^4}{(m_{H_0}+m_N)^2}\, ,
\label{singletdd}
\eea 
where $m_N$ is the nucleon mass, $\l_{HT}$ and $\l_{L}$ are the quartic couplings involved in the DM-Higgs interaction. A recent 
estimate of the Higgs-nucleon coupling ($f$) gives $f=0.32$ \cite{Giedt:2009mr}. We provide below the Feynman diagrams for the spin independent elastic scattering of  DM with nucleon.

\begin{figure}[H]
\centering
\subfigure[]{
\includegraphics[scale=0.420]{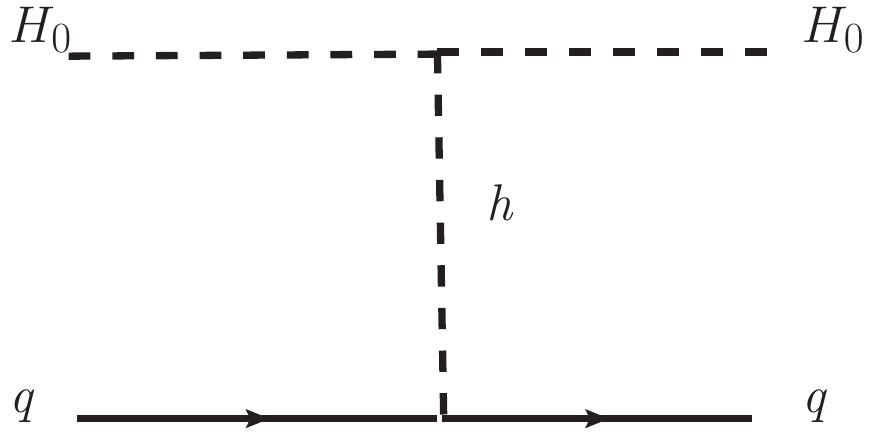}}
\subfigure[]{
\includegraphics[scale=0.450]{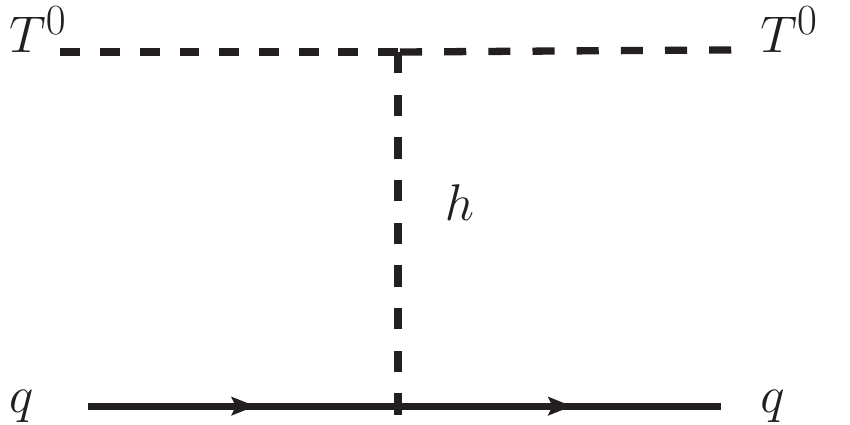}}
\caption{Spin independent elastic scattering of DM-nucleon.}
\label{DD}
\end{figure}
On the other hand, indirect search experiments like Fermi-LAT \cite{MAGIC:2016xys} also offer promising prospects of detecting WIMP DM candidates. Annihilation of DM to SM particles, especially to photons and neutrinos, plays a crucial role here. Since photons and neutrinos are electrically neutral,
they have a higher chance of reaching the detector without getting deflected. The effective indirect detection cross section $\sigma^{\text{ID}}_{\text{i,eff}}$ in a multicomponent DM setup relates to the computed cross section $\sigma^{\text{ID}}_{\text{i}}$ 
as \cite{Bhattacharya:2019fgs,Betancur:2020fdl}
\bea
\sigma^{\text{ID}}_{\text{i,eff}}&=&\bigg(\frac{\Omega_i}{\Omega_{\text{Total}}}\bigg)^2\sigma^{\text{ID}}_{\text{i}}.
\label{ID_expression}
\eea
The exponent 2 in $\big(\frac{\Omega_i}{\Omega_{\text{Total}}}\big)^2$ in case of indirect detection can be explained using the fact that there are two annihilating DM particles in the initial state as opposed to one in case of direct detection. We demand that $\sigma^{\text{ID}}_{\text{i,eff}}$ obey the upper bound from Fermi-LAT for $i = H_0,T_0$.

\subsection{Result}
\label{DM_results}

\begin{figure}[htb!]
\includegraphics[scale=0.38]{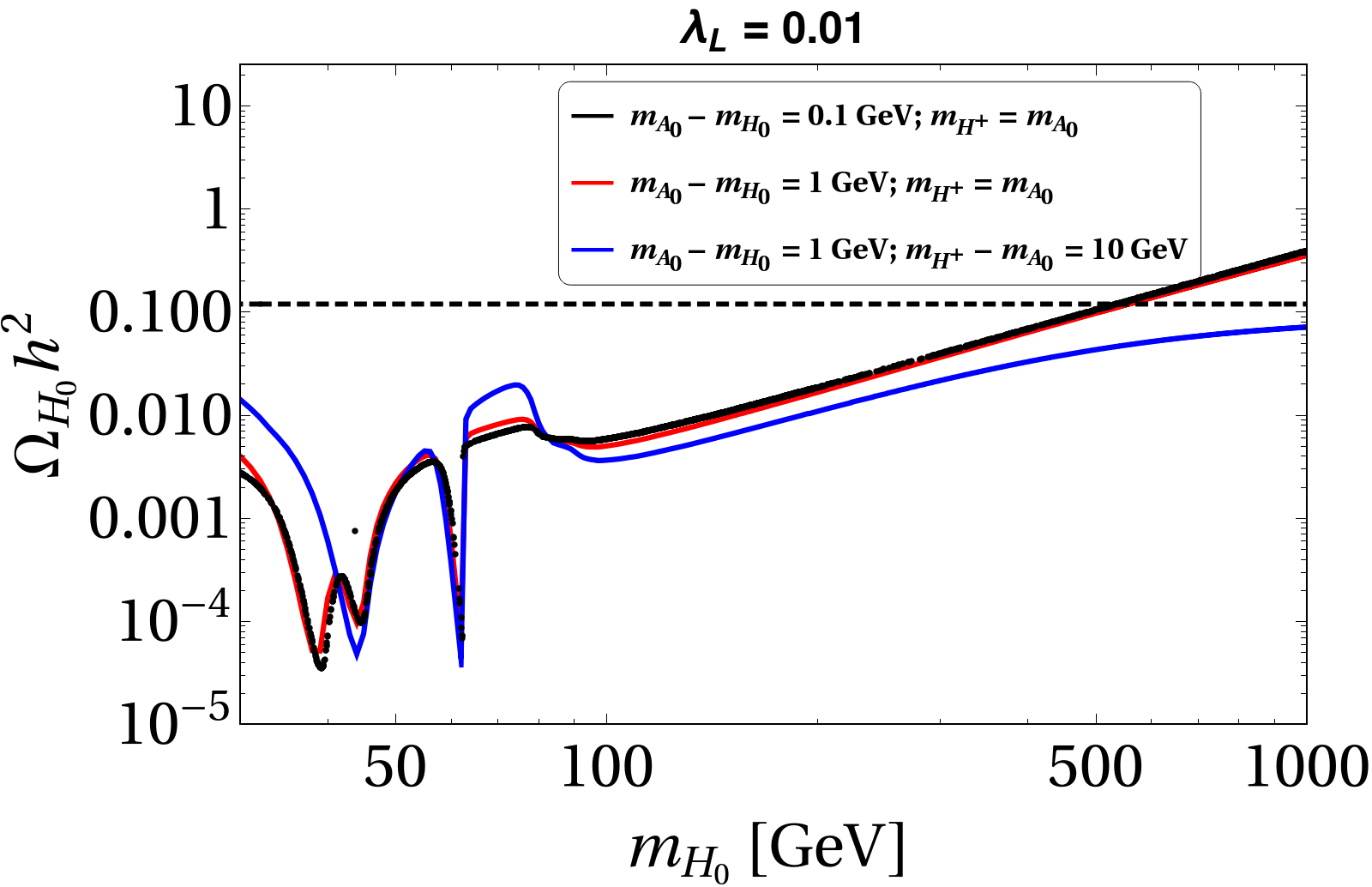}
\includegraphics[scale=0.38]{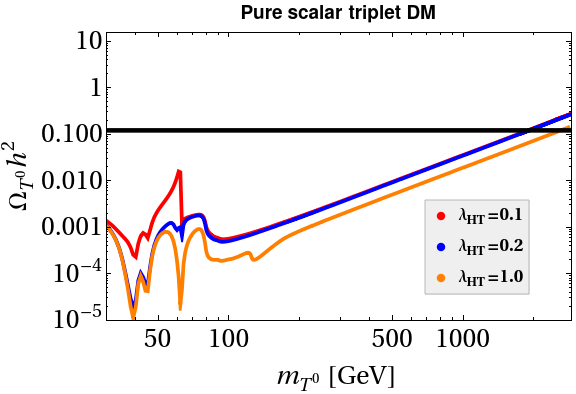}
\caption{Left Panel: The relic density of the inert doublet for $m_{H^+} = m_{A_0}$ and $m_{H^+} =
m_{A_0} + 10$ GeV. Right Panel: Variation of $\Omega_{T_0}h^2$ versus $m_{T_0}$ for different values of $\l_{HT}$.}
\label{pure_DM}
\end{figure}

The phenomenologies of the single component IDM \cite{LopezHonorez:2006gr, Honorez:2010re, Belyaev:2016lok, Choubey:2017hsq, LopezHonorez:2010tb, Ilnicka:2015jba, Arhrib:2013ela, Cao:2007rm, Lundstrom:2008ai, Gustafsson:2012aj, Kalinowski:2018ylg, Bhardwaj:2019mts} as well as ITM \cite{Araki:2011hm, Fischer:2011zz, Fischer:2013hwa, Khan:2016sxm,Jangid:2020qgo,Barman:2021ifu,Bell:2020hnr} are well known. Despite being allowed by the direct search experiments, both fail to predict the observed
for a substantial range of the DM mass. While this under-abundant region extends from 
$M_W$ to 500 GeV for the IDM, the same is in fact larger for the ITM. In this case, the under-abundant region extends till 1.8 TeV of the DM mass (irrespective of the choices of their Higgs portal couplings as their impact on the relic density is sub-dominant) particularly because, apart from its usual annihilations to SM gauge bosons, it must co-annihilate to the gauge bosons with the help of its charged partners $T^{\pm}$ (mass splitting is very small $i.e.~\D m = 166~\text{MeV}$). This in turn makes the effective annihilation cross-section of the triplet DM quite large and hence, leading to under-abundance of the relic density for a wider range of the DM mass 
as compared to the IDM.  We comment on the role of the portal coupling $\l_{HT}$ here. The co(annihilations) are heavily gauge coupling-driven and the dependence on $\l_{HT}$ is subleading. To show this, we plot the triplet relic density for $\l_{HT}$ = 0.1, 0.2, 1 in the right panel of Fig.~\ref{pure_DM}. It is seen that $\Omega_{T_0}$ for $\l_{HT}$ = 0.1, 0.2 differ only slightly away from the $h$-resonance (our region of interest in $m_{T_0} \gtrsim$ 750 GeV). For $\l_{HT}$ = 1, though there appears to be sizeable difference, we can still say that the role of the $h$-mediated annihilation is subleading compared to the gauge-driven annihilatons. This can be understood from the following example. Increasing $\l_{HT}$ from 0.2 to 1 increases the $h$-mediated annihilation cross section 25-fold. However, the relic density for $m_{T_0}$ = 1.8 TeV drops only by a factor $\sim$ 2. This shows that the contribution of the gauge-driven (co)annihilation is way higher. Despite this subleading impact of $\l_{HT}$, a large $\sim$ 1 value for the former can still deplete the relic density by 
$\simeq$ 50$\%$ (\emph{w.r.t.} $\l_{HT}$ = 0.2). This depletion is something we do not aim for and therefore, restrict $\l_{HT}$ in the [0.1,0.2] interval henceforth. And for such a choice, the $T_0 T_0,~T^+ T^- \to H_0 H_0$ conversion is dominated by the $T^{+}-T^{-}-H_0-H_0$ and
$T_0-T_0-H_0-H_0$ contact interactions and not by the s-channel $h$-exchange.

In order to maximise the relic density from the inert doublet, we choose $m_{H^{\pm}}=m_{A_0}$. With $m_{H^{\pm}}\neq m_{A_0}$, this contribution will be accordingly less. We have demonstrated this in the left panel of Fig.\ref{pure_DM} by comparing the relic densities corresponding to $m_{H^+} = m_{A_0}$ and $m_{H^+} = m_{A_0}$ + 10 GeV. The gap thereby made in the total relic density will then have to be compensated by the scalar triplet. Given a heavier $T_0$ tends to predict a higher $\Omega_{T_0}$, the parameter region for $\l_2 \neq \l_3$ would prefer an accordingly heavier $T_0$. This is contrary to our objective of accomplishing a \emph{lighter} $T_0$ (By "lighter", we imply lighter than what is seen in the single component triplet dark matter scenario.).

In the present study, we aim to find out if inter-conversion (which depends on their mass hierarchy though) of one DM species to another can possibly revive such mass regions of IDM and ITM that are known to yield under-abundant thermal relic abundance. As stated before, the relevant parameters that would control the study are $m_{H_0}, m_{A_0},m_{H^{\pm}},\l_L,\l_{HT}$ and $\l_{\Phi T}$. 
For the analysis purpose, we only stick to mass regime $100 ~{\rm{GeV}} \leq m_{H_0}\leq 500~{\rm{GeV}}$ and $100 ~{\rm{GeV}} \leq m_{T_0}\leq 2000~{\rm{GeV}}$ for the doublet and the triplet respectively. As the Higgs portal couplings of both the DM do not play much significant role the in obtaining the correct total relic density for the given choice of the parameter space, we fixed the portal couplings 
$\l_L=0.01$ and $\l_{HT}=0.15$. Even though the role of the portal couplings is subleading in the DM phenomenology, they play a non-trivial role in stabilizing the electroweak vacuum which will be discussed in detail in section \ref{EWSB}. 

\begin{figure}[htb!]
\centering
\subfigure[]{
\includegraphics[scale=0.39]{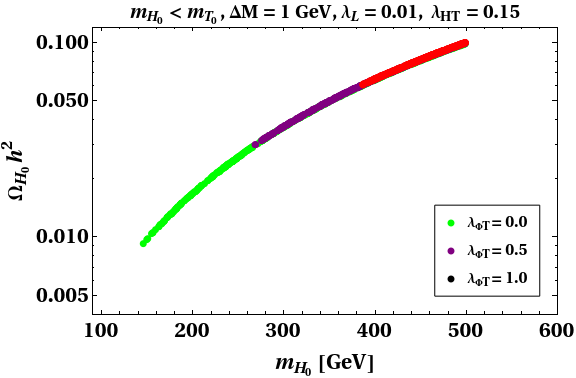}}
\subfigure[]{
\includegraphics[scale=0.39]{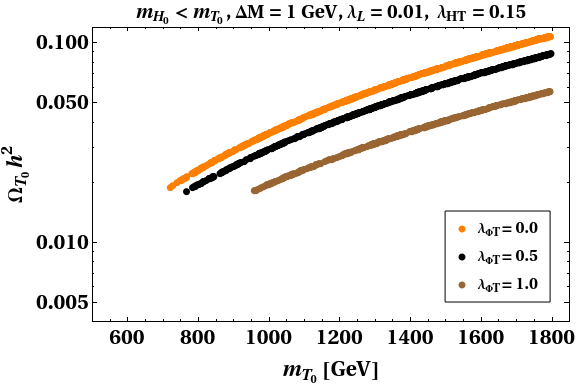}}
\subfigure[]{
\includegraphics[scale=0.39]{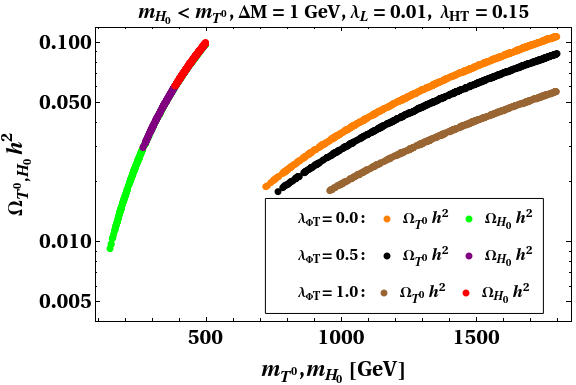}}
\subfigure[]{
\includegraphics[scale=0.39]{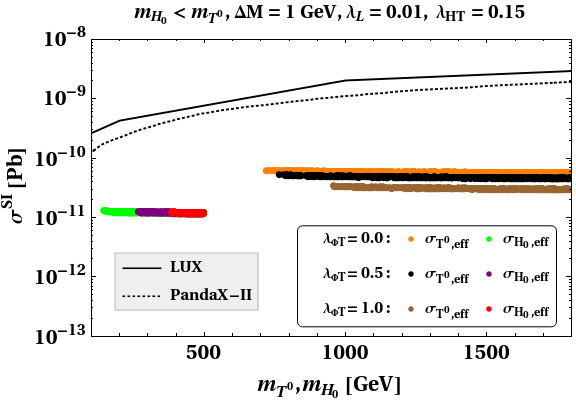}}
\caption{Points that satisfy the correct total DM relic abundance for 
$\lambda_{HT}=0.15$, $\l_L=0.01$, $\Delta M = 1$ GeV and different values of 
$\l_{\Phi T}$ while maintaining $m_{T_0}>m_{H_0}$ in the (a) $m_{H_0}-\Omega_{H_0}h^2$ plane and (b) $m_{T_0}-\Omega_{T_0}h^2$ plane. Panel (c) combines the plots in panels (a) and (b). And panel (d) shows the corresponding Spin independent DM-nucleon scattering cross-sections corresponding to the points in (c). Limits from direct detection experiments are shown in black solid lines (LUX), dotted (PandaX-II) and dashed (XENONnT).}
\label{relic1}
\end{figure}

In Figs.\ref{relic1}(a) and \ref{relic1}(b), we show the variation of the individual relic density contributions, $\Omega_{H_0}h^2$ and $\Omega_{T_0}h^2$, against 
their respective masses, $m_{H_0}$ and $m_{T_0}$, such that the the total relic density ($\Omega_{Total}h^2$) defined in Eq.(\ref{totalrelic}) satisfies the Planck limit \cite{Aghanim:2018eyx}. A combined plot is Fig.\ref{relic1}(c). In the same plot we also show the effect of different choices of the coupling $\l_{\Phi T}$ (involved in DM-DM conversion as shown in Fig.\ref{feynconv}) on the parameter space. Here, we choose three different values of $\l_{\Phi T}$ for illustration, i.e., $\l_{\Phi T}= 0, 0.5$ and $1.0$.  

\noindent The respective variations of the relic densities versus the DM masses with different $\l_{\Phi T}$ are indicated by (i) orange (for $T_0$) and green (for $H_0$) patches for $\l_{\Phi T} = 0$, (ii) black (for $T_0$) and purple (for $H_0$) patches for $\l_{\Phi T} = 0.5$ and (iii) brown (for $T_0$) and red (for $H_0$) patches for $\l_{\Phi T} = 1.0$. Here, for our analysis purpose we also define mass splitting among the inert doublet components as $\Delta M = m_{H^{\pm}}-m_{H_0}= m_{A_0}-m_{H_0}$ and fix its value at 1 GeV. Such a choice of the $\D M$ is motivated from  the fact that the maximum contribution from the single component IDM to relic density can be obtained for such small mass splittings only \cite{LopezHonorez:2006gr, Honorez:2010re, Belyaev:2016lok, Choubey:2017hsq, LopezHonorez:2010tb, Ilnicka:2015jba, Arhrib:2013ela, Cao:2007rm, Lundstrom:2008ai, Gustafsson:2012aj, Kalinowski:2018ylg, Bhardwaj:2019mts}. It is to be noted that the inelastic direct detection processes are ruled out in the IDM for such a mass splitting. In case of the inert
triplet also, the mass gap $m_{T^+} - m_{T_0}$ = 166 MeV is too large to trigger
the $W$-mediated inelastic scattering $T_0 T^+ \to \bar{q} q^\prime$.

We would like to elucidate on Fig.\ref{relic1} a bit.
One must note that $\Omega_{H_0}h^2$ and $\Omega_{T_0}h^2$ in the figure obey
$\Omega_{H_0} h^2 + \Omega_{T_0} h^2$ = Observed relic $\simeq$ 0.12 (the variation within the PLANCK band ignored here for the sake of understanding). For instance, for $\l_{\Phi T}$ = 0.5, $\Omega_{H_0} h^2$ for a particular $m_{H_0}$ can be read from the violet curve. The corresponding $\Omega_{T_0} h^2$ is then approximately equal to 0.12 - $\Omega_{H_0}h^2$ and one can then read the corresponding $m_{T_0}$ value on the X-axis. This implies that the relic densities of the rightmost point on the violet curve and the leftmost point on the black curve add up to $\simeq$ 0.12. This is how Fig.\ref{relic1} should be read\footnote{This approach is based on \cite{Borah:2019aeq}}.

To understand the non-triviality of the conversion coupling in the present set up, we begin with the case where $\l_{\Phi T}=0$. 
Note that with $\l_{\Phi T}=0$, although conversion via process shown in diagram  Fig.\ref{feynconv}(a) does not contribute, 
however the processes 
$T^{+}T^{-},T_0T_0\rightarrow H_{0}H_0$ can still take place due to the non-zero values of $\l_L$ and $\l_{HT}$ 
via the one in Fig.\ref{feynconv}(b). Here, we observe that when $\Omega_{T_0}h^2$ is small (corresponding to lowest point 
of the orange patch), the dominant contribution towards the total relic density comes from $\Omega_{H_0}h^2$ (top most point of the 
green patch which lies below the red patch) so as to satisfy the total relic density with the Planck limit. Similarly a farthest point on the 
orange patch corresponds to the lowest point of the green one.  As an example, for $m_{T_0}= 785 ~\rm{GeV}$ we get $\Omega_{T_0}h^2 = 0.019$ (almost $15 \%$ of the total relic density), rest of the $85\%$ of the total relic density comes from $\Omega_{H_0}h^2$ which corresponds to a single point on the green patch with $m_{H_0}=498$ GeV and $\Omega_{H_0}h^2 =0.098 $. 

Upon turning on the conversion coupling (say $\l_{\Phi T} = 0.5$), 
a shift in the relic densities of both the DM candidates  
is observed (see the black and the purple patches in Fig.\ref{relic1}). The reason behind this is easy to understand.  
When 
the conversion coupling is switched on, the $T_{0}$ starts converting to $H_0$ and hence the relic density of the $T_0$ decreases 
whereas we observe an upward shift for the relic density $\Omega_{H_0}$ (the purple patch). A similar behavior is observed for $\l_{\Phi T} = 1.0$   where the relic density of $T_0$ is further decreased and the relic density of $H_0$ is increased, we notice that the red patch has now became much smaller. This is because the maximum contribution towards the total relic density from $T_{0}$ can at most be $53\%$ (with $m_{T_0}=1790$ GeV, $\Omega_{T_0}h^2=0.060$)  which in turn requires $\Omega_{H_0}h^2=0.056$ (rest of the $47\%$ contribution towards the total relic density) for $m_{H_0}=387$ GeV leading to a shrinking in the red patch. 
In Fig.\ref{relic1}(d), we plot the effective direct detection cross-section of both the DM with respect to their respective masses and compare it with the experimental results obtained from LUX \cite{Akerib:2016vxi}, PandaX-II \cite{Tan:2016zwf,Cui:2017nnn} and XENONnT~\cite{XENON:2020kmp} for different values of $\l_{\Phi T}$ (similar to Fig.\ref{relic1}(a)).  Here it is interesting to point out that, although the current setup is allowed from the constraints coming from present direct search experiments like LUX \cite{Akerib:2016vxi} and PandaX-II \cite{Tan:2016zwf,Cui:2017nnn}, the parameter space of the setup can come in tension with the XENONnT projections. However, this is actually a positive finding since this renders the model testable and hence falsifiable in a future experiment. Considering the Fig.\ref{relic1}(a) and the bounds from the present direct search results from Fig.\ref{relic1}(d) together, we can conclude that the parameter space under consideration is allowed from both the relic density as well as the direct detection constraints. For better understanding, we also tabulate the result discussed above in Table \ref{tab_new}  for three different choices of the conversion couplings $\l_{\Phi T}=0.0,0.5$ and 1.0.\\

\begin{table}
\centering
\begin{tabular}{|c|c|c|c|c|}
\hline
$\l_{\Phi T}$ & $m_{H_0}$ [GeV] & $m_{T_0}$ [GeV] & $~\Omega_{H_0}h^2$ & $~\Omega_{T_0}h^2$  \\
\hline
0.0 & 500 & 730 & 0.0981 & 0.0192  \\
    & 147 & 1799 & 0.0092 & 0.1070    \\
\hline
0.5 & 500 & 770 & 0.0988 & 0.0180 \\ 
    & 280 & 1799 & 0.0320 & 0.0875 \\
\hline
1.0 & 500 & 1000 & 0.0993 & 0.0194 \\ 
    & 388 & 1799 & 0.0606 & 0.0568 \\
\hline
\end{tabular}
\caption{Table showing the relic densities for certain sample mass values for different
$\l_{\Phi T}$. We have taken $\l_{L}=0.01$ and $\l_{HT}=0.15$. }
\label{tab_new}
\end{table}

\begin{figure}
\centering
\subfigure[]{
\includegraphics[scale=0.39]{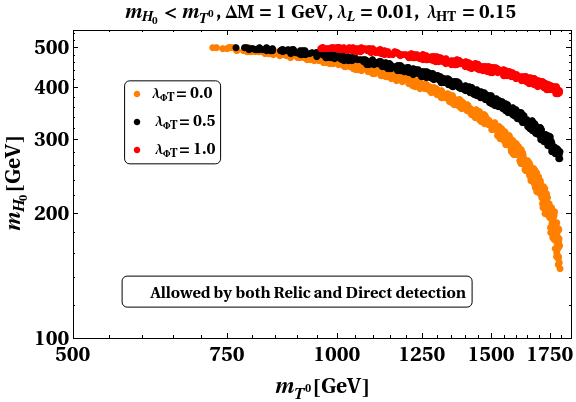}}
\subfigure[]{
\includegraphics[scale=0.39]{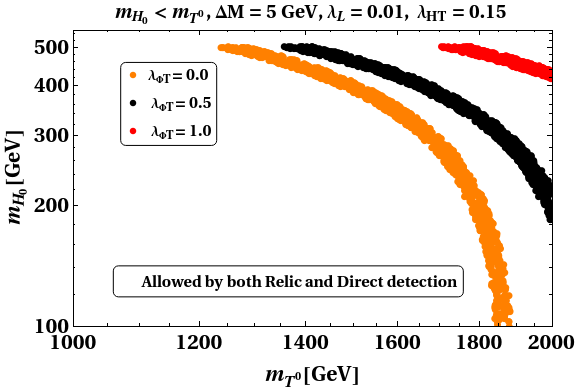}}
\caption{All the points which satisfy the correct total DM relic abundance and are also allowed by direct detection for different values of $\l_{\Phi T}$ while maintaining $ m_{T_0}>m_{H_0} $ for  $\lambda_{HT}=0.15$ and (a) $\Delta M = 1$ GeV, (b) $\Delta M = 5$ GeV in $m_{H_0}-m_{T_0}$ plane. }
\label{mHmT1}
\end{figure}

Finally, in Fig.\ref{mHmT1}, we incorporate parameter region plots in the $m_{H_0}-m_{T_0}$ plane for $\Delta M$ = 1 GeV (left panel), 
5 GeV (right panel). All the points in this case are 
allowed by the relic density and direct detection constraints. The effect of the conversion coupling discussed in Fig.\ref{relic1}(a) 
becomes prominent in Fig.\ref{mHmT1}.
Looking at the left panel of Fig.\ref{mHmT1}, one may notice that 
almost the 
entire desert mass regime of the single component IDM and $m_{T_0}$ in the range $700~{\rm{GeV}} \leq m_{T_0}\leq 2 ~{\rm{TeV}}$ 
(The relic density of a $Y$=0 triplet being underabundant for $m_{T_0} <$ 1.8 TeV) now becomes allowed in the two component 
set up, thanks to DM-DM conversion. However, for the  higher value $\Delta M$ = 5 GeV  
we observe in the right panel that $m_{T_0}$ shifts towards the heavier side. This happens because with 
the increase in the mass splitting among the inert doublet components, the contribution of $\Phi$ to the total relic density decreases and hence in order for total relic 
density to satisfy the Planck limit, the triplet contribution has to increase which can only result from an accordingly larger 
mass of the triplet DM.
 \begin{figure}[H]
\centering
\subfigure[]{
\includegraphics[scale=0.45]{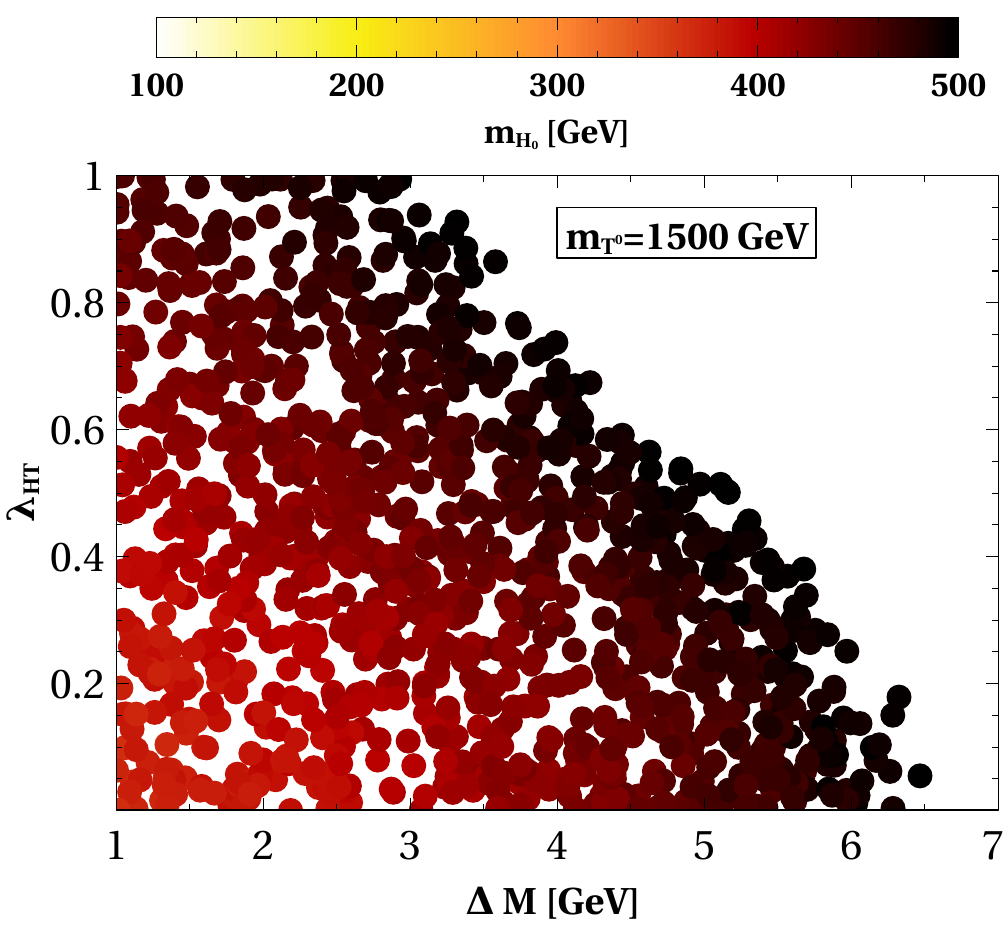}}
\caption{Variation of Triplet-Higgs portal coupling ($\l_{HT}$) with inert doublet mass splitting ($\D M$) for a fixed value of $m_{T_0}=1500$ GeV. Colour bar shows the variation with $m_{H_0}$.}
\label{lHT_delM}
\end{figure}

Although the effect of the Higgs portal coupling $\l_{HT}$ is subleading in the present scenario, a small effect of its variation can be observed in the parameter space when we vary the mass splitting among the inert doublet components. 
We try to demonstrate this subleading behaviour via a heat plot in Fig.\ref{lHT_delM}. For a fixed $m_{T_0}$, the annihilation rate of the scalar triplet increases upon increasing 
$\l_{HT}$ thereby causing $\Omega_{T_0}h^2$ to drop, albeit mildly. Since the total relic density is demanded to lie in the Planck band, 
$\Omega_{H_0}h^2$ must accordingly increase to fill the gap.  
And that is possible only for a higher inert doublet DM mass. And since, the larger the $m_{H_0}$, the darker is the shade, one expects a darkening of the points as one moves up the $\l_{HT}$ axis. This is exactly what is observed in Fig.\ref{lHT_delM} where a gradual darkening is seen as one moves up from the bottom left to the top left corner. On the other hand, as $\D M$ increases, $\Omega_{H_0}h^2$ tends to decrease for a given $m_{H_0}$. If $\l_{HT}$ is held fixed, one then has to then increase $m_{H_0}$ to maintain the original $\Omega_{H_0}h^2$. A darkening of the points is then expected as one moves towards the right of the $\D M$ axis. An inspection of the figure corroborates this when such a darkening is immediately seen.

Prior to ending this section, we comment on the consequences of an $m_{H_0} > m_{T_0}$ hierarchy. Here the processes responsible for DM conversion will now be reversed and the inert
doublet is expected to be responsible for the production of the second dark matter $T_0$
i.e. $H_0 H_0 \to T_0 T_0$. As mentioned before, a sub-TeV scalar triplet always leads to an under
abundant relic due to high co(annihilation) rates to gauge boson final states. Given that we
restrict $m_{H_0} < $ 500 GeV in this study, demanding $m_{H_0} > m_{T_0}$ at the same time implies that
even the $H_0 H_0 \to T_0 T_0$ conversion does not suffice to generate the observed relic. Therefore, the
 $m_{H_0} > m_{T_0}$ hierarchy is not appealing from the two component DM perspective and we shall not consider it any further.

\section{EW vacuum stability and combined analysis}
\label{EWSB}

This section discusses the role of the additional scalar multiplets in (meta)stabilising the EW vacuum. For high field values, i.e., $h > > v$, the RG improved effective potential for this scenario can be expressed as \cite{Degrassi:2012ry}
\bea
V^{\rm{eff}}_H &\simeq& \frac{\l^{\rm{eff}}_H(h)}{4}h^4, 
\label{effpot}   
\eea        
\noindent where $\l^{\rm{eff}}_H(\m)=\l_H(\m)+\l^{\rm{SM,eff}}_H(\m)+\l^{\Phi,\rm{eff}}_H(\m)+\l^{\rm{T,eff}}_H(\m)$. Here, $\l^{\rm{SM,eff}}_H(\m)$ is the contribution coming from the SM fields to $\l_H$ whereas $\l^{\rm{\Phi,eff}}_H(\m)$ and $\l^{\rm{T,eff}}_H(\m)$ are the contributions from $\Phi$ and 
$T$ respectively. The EW boundary scale from which the couplings start evolving is chosen to be the $t$-quark pole mass $M_t$ = 173.34 GeV. All the running couplings are to be evaluated at a scale $\mu = h$ in Eq.(\ref{effpot}).
One derives the following.
\bea
\l^{\rm{\Phi,eff}}_H(\mu) &=& e^{4\G(\m)}\frac{1}{16\pi^2} \bigg[2\frac{\l_{1}^2}{4}\bigg(\rm{ln}\frac{\l_{1}}{2}-\frac{3}{2}\bigg)+\frac{(\l_{1}+\l_2+\l_3)^2}{4}\bigg(\rm{ln}\frac{\l_{1}+\l_2+\l_3}{2}-\frac{3}{2}\bigg)\nonumber \\
&&+\frac{(\l_{1}+\l_2-\l_3)^2}{4}\bigg(\rm{ln}\frac{\l_{1}+\l_2-\l_3}{2}-\frac{3}{2}\bigg)\bigg],\\
\l^{\rm{T,eff}}_H(\m)&=& e^{4\G(\m)} \frac{1}{16\pi^2}\bigg[\frac{3\l^2_{HT}}{4}\bigg(\text{ln}\frac{\l_{HT}}{2}-\frac{3}{2}\bigg)\bigg] .  
\eea 
\label{eff_lamH} 
Here, $\G(\mu)=\int_{M_t}^{\mu}\gamma(\mu^\prime)~\rm{d~ln(\m^\prime)}$ and $\gamma(\m)$ denotes the anomalous dimension of the Higgs field \cite{Buttazzo:2013uya}. By virtue of such quantum effects, a second minima can show up at high energy scales. The condition 
$\l^{\text{eff}}_H(\mu) > 0$ ensures that the EW minimum is deeper than the second minimum, that is, a \emph{stable}
EW vacuum. On the other hand, 
$\l^{\text{eff}}_H(\mu) < 0$ implies that the second minimum is deeper. The fate of the EW vacuum in this case is decided by computing the probability of tunnelling to the second vacuum. The expression for the tunnelling probability is given by
\bea
\mathcal{P}_T &=& (\mu_B T_U)^4 e^{-\frac{8\pi^2}{3 \l_H(\mu_B)}}.
\label{prob}
\eea
In Eq.(\ref{prob}), $T_U$ is the age of the universe and $\mu_B$ denotes the scale at which the tunneling probability is
maximized, determined from $\beta_{\l_H}(\mu_B)$ = 0. The EW vacuum is \emph{metastable} if the tunnelling lifetime is greater than the universe's age. With this, one obtains the following criterion on $\l^{\text{eff}}_H (\mu)$:
\bea
\l^{\text{eff}}_H (\mu) &>& \frac{-0.065}{1 - \text{ln}
(v/\mu_B)}.
\eea
The following boundary values are then taken for the SM Yukawa and gauge couplings~\cite{Buttazzo:2013uya}\footnote{Heaviness of the IDM and ITM masses implies that their possible threshold contributions to the gauge and Yukawa couplings are negligible.}. 
\besub
\bea
y_{t}(M_t) &=& 0.93690 + 0.00556 \times (M_t - 173.34) - 0.00042 \times (\a_s(M_Z) - 0.1184)/0.0007,\nonumber\\ \\
g_{1}(M_t) &=& 0.35830 + 0.00011 \times (M_t - 173.34) - 0.00020 \times (M_W - 80.384)/0.014, \\
g_{2}(M_t) &=& 0.64779 + 0.00004 \times (M_t - 173.34) + 0.00011 \times (M_W - 80.384)/0.014, \\
g_{3}(M_t) &=& 1.1666 - 0.00046 \times (M_t - 173.34) + 0.00314 \times (\a_s(M_Z) - 0.1184)/0.0004.
\eea
\eesub 
We take $M_W$ = 80.384 GeV and $\a_s(M_Z)$ = 0.1184. The input values of $\l_1,\l_2$ and $\l_3$ are determined
using Eqs.~(\ref{l1})-(\ref{l3}). 

\begin{figure}[H]
\centering
\includegraphics[scale=0.43]{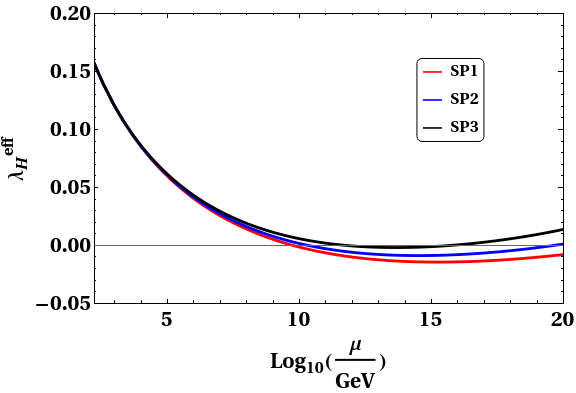}~~~
\includegraphics[scale=0.43]{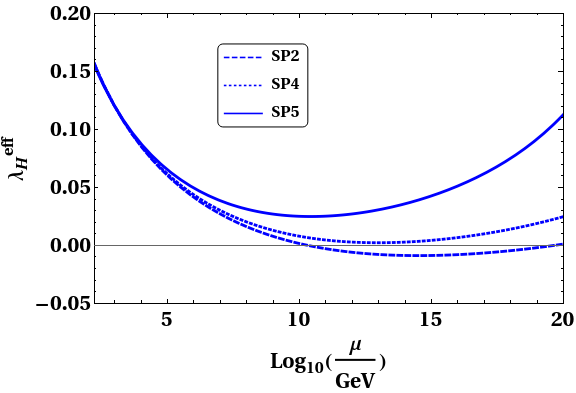} 
\caption{RG evolution of $\l_{H}^{\text{eff}}$ for the SPs. In the left panel for SP1, SP2 and SP3 the values for $\{\l_1,~\l_2,~\l_3\}= \{0.046,-0.013-0.013\}$ while in the right panel for SP4 $\{\l_1,~\l_2,~\l_3\}=\{0.153,~-0.066,~-0.066\}$ and for SP5  $\{\l_1,~\l_2,~\l_3\}=\{0.287,~-0.134,~-0.134\}$.}
\label{f:running}
\end{figure}

The expression for $\beta_{\l_H}$ in Eq.(\ref{running_scalar}a) tells us that the quartic couplings the vacuum instability (or metastability) scale is sensitive to are $\l_{1-3}$ and $\l_{HT}$. The sample points (SPs) listed in Table~\ref{Tab:BP1} demonstrate this sensitivity.
\begin{table}
\centering
\begin{tabular}{|c|c|c|c|c|c|c|c|c|c|c|c|}
\hline
SP & $m_{H_0}$ & $m_{T_0}$ & $\l_{HT}$ & $\D M$ & $\mu_{\gamma \gamma}$ & $\Omega_{T_0}h^2$ &$\Omega_{H_0}h^2$&$\sigma_{T_0,eff}$ ($pb$)&$\sigma_{H_0,eff}$($pb$)&$\sigma^{\text{ID}}_{T_0,eff}$($cm^3/s$)&$\sigma^{\text{ID}}_{H_0,eff}$($cm^3/s$)\\ \hline
SP1 &$400$ & $1426$ & 0.1 & $1$ & 0.999 & $0.056$ &$0.064$ &$2.01\times10^{-11}$ &$1.2\times10^{-11} $&$7.1\times10^{-28} $&$1.2\times10^{-26} $\\
\hline
SP2 &$400$ & $1426$ & 0.15 & $1$ & 0.999 & $0.056$ &$0.064$ &$4.5\times10^{-11}$ &$1.2\times10^{-11} $&$7.1\times10^{-28} $&$1.2\times10^{-26} $\\
\hline
SP3 &$400$ & $1426$ & 0.2 & $1$ & 0.999 & $0.056$ &$0.064$ &$8.1\times10^{-11}$ &$1.2\times10^{-11} $&$7.0\times10^{-28} $&$1.2\times10^{-26} $\\
\hline
SP4 &$400$ & $1670$ & 0.15 & $5$ & 0.997 & $0.076$ & $0.044$ &$4.5\times10^{-11}$ &$8.0\times10^{-12} $&$2.1\times10^{-27} $&$5.3\times10^{-27} $\\
\hline
SP5 &$400$ & $1856$ & 0.15 & $10$ & 0.994 & $0.093$ & $0.025$ & $4.5\times10^{-11}$ &$4.6\times10^{-12}$&$5.9\times10^{-27} $&$1.7\times10^{-27} $\\
\hline
\end{tabular}
\caption{Sample points predicting a total relic density within the Planck band that are also allowed by the theoretical, direct detection, indirect detection and diphoton signal strength constraints. All masses and mass-splittings are in GeV. 
}
\label{Tab:BP1}
\end{table}
We choose $\l_L$ = 0.01, $\l_{\Phi T} = 0.5$, $\l_{\phi}=0.001$ and $\l_{T}=0.001$ throughout the analysis using RG.
It is noted that the SP1-3 differ only in their values of $\l_{HT}$. Since (co)annihilation in the triplet sector is heavily driven by gauge interactions, tuning $\l_{HT}$ in the interval [0.1,0.2] changes 
$\Omega_{T_0}$ only slightly. The corresponding RG trajectories of 
$\l_{H}^{\text{eff}}$ is shown in the left panel of Fig.\ref{f:running}. It is seen that though $\l_{H}^{\text{eff}}$ turns negative in each case, it remains within the metastable band. In addition, the larger the input value of 
$\l_{HT}$, the higher is the scale at which $\l_{H}^{\text{eff}}$ turns negative. As mentioned before, this is solely due to the presence of the $\mathcal{O}(\l_{HT}^2)$ term in $\beta_{\l_H}$. While SP4 and SP5 are primarily characterized by their values of 
$\Delta M$, the $m_{T_0}$ values are also different for each. This is because a decrease in $\Omega_{H_0}$ that inevitably occurs with an increasing $\Delta M$ in these sample points is counterbalanced by an increased contribution from the triplet, something in turn achieved by appropriately raising $m_{T_0}$. Now, with $\l_L$ fixed, increasing $\Delta M$ accordingly increases the magnitudes of 
$\l_{1-3}$ at the EW scale. This in turn generates an upward push to the RG trajectory of $\l_H^{\text{eff}}$ via the $\l_{1-3}$-dependent terms in 
$\beta_{\l_H}$. Since SP2, SP4 and SP5 feature different $\Delta M$ values for the same $\l_{HT}$, we show the corresponding RG evolution curves in the right panel of Fig.\ref{f:running} in order to confirm the impact of changing $\Delta M$. A metastable EW vacuum is identified for $\Delta M$ = 1 GeV. Increasing the same to $\Delta M$ = 5 GeV and 10 GeV stabilises the same up to the Planck scale. 

\begin{figure}
\centering
\includegraphics[scale=0.41]{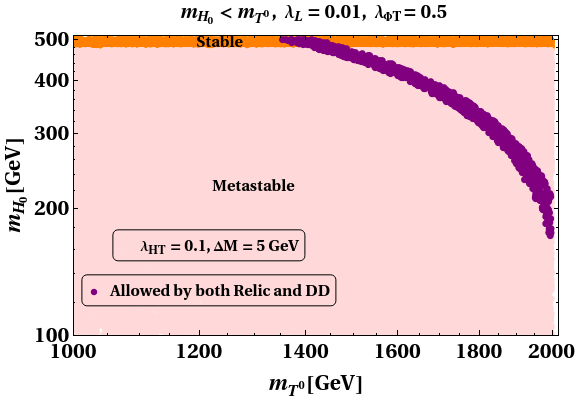}~~~
\includegraphics[scale=0.41]{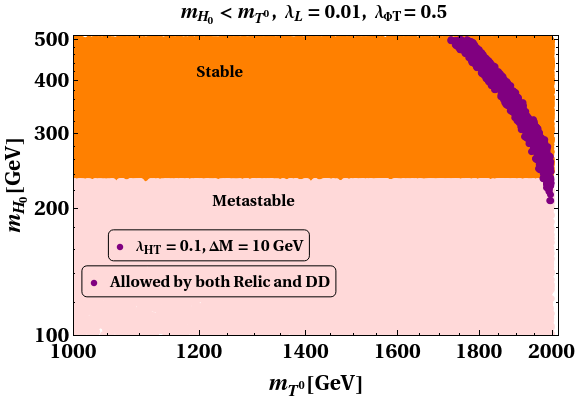} \\
\includegraphics[scale=0.41]{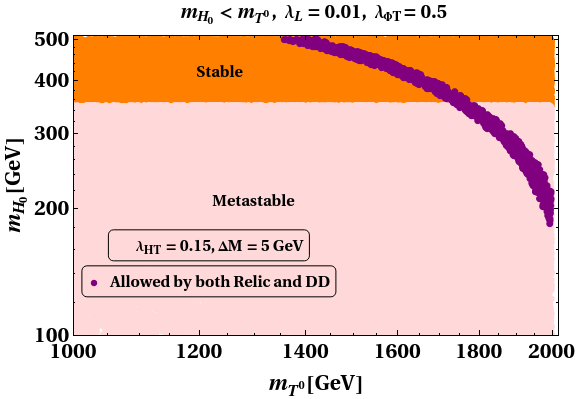}~~~
\includegraphics[scale=0.41]{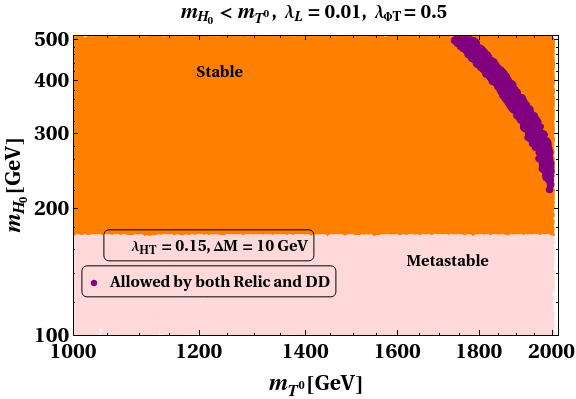}
\caption{Parameter region compatible with (meta)stable vacuum and the DM observables in the $m_{T_0}-m_{H_0}$ plane. The orange (pink) bands correspond to stability (metastability). The blue region is allowed by the DM constraints.}
\label{f:combined}
\end{figure}

The bands corresponding to metastable and stable EW vacuum sketched in the $m_{T_0}-m_{H_0}$ plane are shown in Fig.\ref{f:combined}. Also, the parameter region predicting the requisite relic density and compatible with the direct detection constraints is overlaid on the same. Upon inspection, it is seen that the parameter plane mostly favours metastability over stability for $\l_{HT}$ = 0.1 and $\Delta M$ = 5 GeV. And this is attributed to the EW scale values of $\l_1,\l_2$
$\l_3$ and $\l_{HT}$ that are not sizeable enough to stabilise the vacuum till $M_{\text{Pl}}$ for most part of the $m_{T_0}-m_{H_0}$ plane. In fact, a stable vacuum is ruled out for $m_{H_0} \lesssim$ 480 GeV. With increasing 
$\l_{HT}$ to 0.15, the band corresponding to stability expands to include $m_{H_0} \gtrsim 360$ GeV. This is expected for the following reason. For a higher $\l_{HT}$, accordingly smaller $|\l_1|,|\l_2|$ and $|\l_3|$ suffice to ensure a stable vacuum up to $M_{\text{Pl}}$. And with $\l_L$ and $\Delta M$ fixed, smaller $|\l_1|,|\l_2|$ and $|\l_3|$ imply a smaller $m_{H_0}$. We also remark that the parameter region compatible with the DM constraints changes only slightly with this change in $\l_{HT}$ since the (co)annihilations of the triplet scalars are mostly driven by the gauge interactions. Demanding the requisite relic abundance rules out $m_{T_0} \lesssim 1.35$ TeV for 
$\Delta M$ = 5 GeV. 

As illustrated before in Fig.\ref{f:running}, increasing $\Delta M$ while keeping $\l_L$ unchanged aids vacuum stability by increasing the magnitudes of $\l_1,\l_2$ and $\l_3$. In other words, with a higher 
$\Delta M$, the requisite magnitudes of these quartic couplings at the input scale required to stabilise the EW vacuum up to the Planck scale can be achieved for an accordingly lower $m_{H_0}$. This is concurred by the plot with $\l_{HT}$ = 0.1 and $\Delta M$ = 10 GeV in which case the stability band includes $m_{H_0} \gtrsim 240$ GeV. In addition, the DM-compatible region also changes appreciably with respect to $\Delta M$ = 5 GeV. With now a higher mass-splitting in the inert doublet sector, $\Omega_{H_0}$ diminishes. Heavier triplet scalars are needed to fill up the deficit in relic density compared to what occurs for $\Delta M$ = 5 GeV.
This is the reason why the DM-compatible region shifts towards right in the $m_{T_0}-m_{H_0}$ plane when switching from $\Delta M$ = 5 GeV to 10 GeV. The observed relic abundance in fact obviates $m_{T_0} \lesssim$ 1.7 TeV for $\Delta M$ = 10 GeV. Once again, we would like to contrast this finding with the standalone IDM. In the present IDM+ITM setup, a smaller $\Delta M$ is required to stabilise the vacuum till the Planck scale compared to what would be required in case of the standalone IDM. This is expected on grounds of additional bosonic contribution coming from the triplet in case of the IDM+ITM. To cite an example parameter point, for $M_H$ = 250 GeV, $\l_L$ = 0.01, a stable EW vacuum within the standalone IDM mandates the higher mass gap $\Delta M \gtrsim 14$ GeV. And we reiterate that the relic density also remains underabundant for the same. Therefore, the introduction of an additional inert triplet significantly modifies the analyses of both DM phenomenology and EW vacuum stability. While the IDM desert region now becomes compatible with the observed relic density, it also predicts a stable EW vacuum all the way up to the Planck scale for a lower $\Delta M$.

\section{Conclusions}
\label{Conclusions}

Despite its popularity, the standalone inert scalar doublet fails to account for the observed thermal relic abundance in the desert region, i.e., 100 GeV  $<M_{DM}<$  500 GeV. Similarly, a $Y$ = 0 inert triplet is seen to yield only under-abundant relic density for $M_{DM}<$  1.8 TeV. 
In this work, we extend the scalar sector of the SM by both the aforementioned multiplets and impose a $Z_2 \times Z_2^\prime$ symmetry so that the scalar doublet and triplet constitute two different dark sectors. A two component DM scenario is consequently realised with the neutral CP-even scalar from each multiplet as a DM candidate. The quartic coupling 
$\l_{\Phi T}$, that controls the rate of DM-DM conversion, emerges as a crucial parameter. Also important in the context of DM phenomenology is the mass-splitting between the inert scalars on which the corresponding relic density is sensitive to. For appropriate choice of the relevant model parameters, we demonstrate how DM-DM conversion is instrumental in generating the observed relic densities for the doublet scalar mass in the desert region, and, a sub-TeV triplet scalar. Moreover, this observation is found to be consistent with the latest DD bound. 

We also compute the one-loop RG equations for the present framework and subsequently discuss (meta)stability of the EW vacuum. Demanding a stable vacuum up to the Planck scale in addition to the requisite relic density further restricts the mass regions of interest. For instance,
a stable vacuum till the Planck scale disfavours sub-TeV triplet DM for the mass-splitting among the doublet scalars not exceeding 10 GeV. In all, we show that DM-DM conversion in the present multi-component DM model can lead to an explanation of the observed relic density in specific mass regions that are otherwise known to yield under-abundance in the corresponding single component cases. That the aforementioned observation is compatible with a stable vacuum up to the Planck scale, is a major upshot of this analysis.  
Lastly, we add that the present scenario also bears
an interesting discovery potential at LHC. It is in fact more prospective to probe a \emph{lighter} charged triplet scalar via the disappearing charge track at the detector. This direction warrants a separate investigation in the near future.

\section{Acknowledgements}

NC is
financially supported by IISc (Indian Institute of Science, Bangalore, India) through the C.V. Raman
postdoctoral fellowship. NC also acknowledges support from DST, India, under grant number IFA19-
PH237 (INSPIRE Faculty Award).  

\appendix
\section{Beta Functions}
The one-loop beta functions for the quartic couplings can be split as
 $\beta_{\l_i} = \beta^{S}_{\l_i} + \beta^{F}_{\l_i}+ \beta^{G}_{\l_i}$. Then
\besub
\bea
16 \pi^2 \beta^S_{\l_H} &=& 24 \l_H^2 + 2 \l_1^2 + 2 \l_1 \l_2 + \l_2^2 + \l_3^2 + \frac{3}{2}\l_{HT}^2, \\
16 \pi^2 \beta^S_{\l_\phi} &=& 24 \l_\phi^2 + 2 \l_1^2 + 2 \l_1 \l_2 + \l_2^2 + \l_3^2 + \frac{3}{2}\l_{\Phi T}^2, \\
16 \pi^2 \beta^S_{\l_T} &=& \frac{11}{3}\l_T^2 + 12 \l_{H T}^2
+ 12 \l_{\Phi T}^2, \\
16 \pi^2 \beta^S_{\l_1} &=& 4 \l_1^2 + 2 \l_2^2 + 
2 \l_3^2
+ 12 \l_1 \l_H + 4 \l_2 \l_H + 12 \l_1 \l_\phi \nonumber \\&+& 4 \l_2 \l_\phi + 3 \l_{H T} \l_{\Phi T}, \\
16 \pi^2 \beta^S_{\l_2} &=& 4 \l_2^2 + 8 \l_3^2 + 8 \l_1 \l_2
+ 4 \l_2 \l_H + 4 \l_2 \l_\phi, \\
16 \pi^2 \beta^S_{\l_3} &=& 8 \l_1 \l_3 + 12 \l_2 \l_3 + 4 \l_3 \l_H + 4 \l_3 \l_\phi, \\
16 \pi^2 \beta^S_{\l_{HT}} &=&  4 \l_{HT}^2 + 12 \l_H \l_{HT} 
+ 4 \l_1 \l_{\Phi T} + 2 \l_2 \l_{\Phi T} \nonumber \\&+& \frac{5}{3}
\l_{H T}\l_T, \\
16 \pi^2 \beta^S_{\l_{\Phi T}} &=&  4 \l_{\Phi T}^2 + 12 \l_\phi \l_{\Phi T} 
+ 4 \l_1 \l_{H T} + 2 \l_2 \l_{H T} \nonumber \\&+& \frac{5}{3}
\l_{\Phi T}\l_T.
\eea
\label{running_scalar}
\eesub

\besub
\bea
16 \pi^2 \beta^F_{\l_H} &=& 12 \l_H y_t^2 - 6 y_t^4, \\
16 \pi^2 \beta^F_{\l_\phi} &=& 0, \\
16 \pi^2 \beta^F_{\l_T} &=& 0, \\
16 \pi^2 \beta^F_{\l_1} &=& 6 \l_1 y_t^2, \\
16 \pi^2 \beta^F_{\l_2} &=& 6 \l_2 y_t^2, \\
16 \pi^2 \beta^F_{\l_3} &=& 6 \l_3 y_t^2, \\
16 \pi^2 \beta^F_{\l_{H T}} &=& 6 \l_{H T} y_t^2, \\
16 \pi^2 \beta^F_{\l_{\Phi T}} &=& 0.
\eea
\eesub

\besub
\bea
16 \pi^2 \beta^G_{\l_H} &=& -3\l_H(g_1^2 + 3 g_2^2)
+ \frac{3}{8}(g_1^4 + 3 g_2^4 \nonumber \\&+& 2 g_1^2 g_2^2), \\
16 \pi^2 \beta^G_{\l_\phi} &=& -3\l_\phi(g_1^2
 + 3 g_2^2)
+ \frac{3}{8}(g_1^4 + 3 g_2^4 \nonumber \\&+& 2 g_1^2 g_2^2), \\
16 \pi^2 \beta^G_{\l_T} &=& -24 \l_T g_2^2 + 72 g_2^4, \\
16 \pi^2 \beta^G_{\l_1} &=& -3\l_1(g_1^2
 + 3 g_2^2)
+ \frac{3}{4}(g_1^4 + 3 g_2^4 \nonumber \\&-& 2 g_1^2 g_2^2), \\
16 \pi^2 \beta^G_{\l_2} &=& -3\l_2(g_1^2 + 3 g_2^2)
+ 3 g_1^2 g_2^2, \\
16 \pi^2 \beta^G_{\l_3} &=& -3\l_3(g_1^2 + 3 g_2^2), \\
16 \pi^2 \beta^G_{\l_{H T}} &=& -\l_{H T}
\Big( \frac{3}{2}g_1^2 + \frac{33}{2} g_2^2 \Big)
 + 6 g_2^4,
 \\
16 \pi^2 \beta^G_{\l_{\Phi T}} &=& -\l_{\Phi T}
\Big( \frac{3}{2}g_1^2 + \frac{33}{2} g_2^2 \Big)
 + 6 g_2^4. 
\eea
\eesub

The $t$-Yukawa evolves according to
\bea
16\pi^2 \beta_{y_t} &=& \frac{9}{2}y_t^3
 - y_t\bigg(\frac{17}{12}g_1^2 + \frac{9}{4}g_2^2
 + 8 g_3^2\bigg).
\eea

Finally, the gauge couplings have the following beta functions.
\besub
\bea
16 \pi^2 \beta_{g_1} &=& 7 g_1^3, \\
16 \pi^2 \beta_{g_2} &=& -\frac{8}{3} g_2^3, \\
16 \pi^2 \beta_{g_3} &=& 16 \pi^2 \beta^{\text{SM}}_{g_3}
= -7 g_3^3.
\eea
\eesub

\bibliographystyle{utphys}
\bibliography{ref.bib}


\end{document}